\documentclass[journal,a4paper]{IEEEtran}
\usepackage[utf8]{inputenc}
\usepackage{amsmath}
\usepackage{mathtools}
\usepackage{graphicx}
\usepackage{float}
\usepackage{caption}
\usepackage{cite}
\usepackage{blindtext}
\usepackage{xcolor}
\usepackage{hyperref}
\usepackage{caption}
\usepackage[utf8]{inputenc}
\newcommand{\rom}[1]{\uppercase\expandafter{\romannumeral #1\relax}}
\usepackage{amssymb}

\begin{document}

\title{Modelling the delayed nonlinear fiber response in coherent optical communications}
%
%

\author{Daniel~Semrau,~\IEEEmembership{Student Member,~IEEE,} Eric~Sillekens,~\IEEEmembership{Student Member,~IEEE,} Robert~I.~Killey,~\IEEEmembership{Senior Member,~IEEE,} and~Polina~Bayvel,~\IEEEmembership{Fellow,~IEEE,~Fellow,~OSA}
\thanks{This work was supported by a UK EPSRC programme grant TRANSNET (EP/R035342/1).}
\thanks{D. Semrau, E. Sillekens, R. I. Killey and P. Bayvel are with the Optical Networks Group, University College London, London
WC1E 7JE, U.K. (e-mail: \{uceedfs; e.sillekens; r.killey; p.bayvel\}@ucl.ac.uk.)}
}

\maketitle

\markboth{\today}%
{}

\begin{abstract}
Fiber nonlinearities, that lead to nonlinear signal interference (NLI), are typically regarded as an instantaneous material response with respect to the optical field. However, in addition to an instantaneous part, the nonlinear fiber response consists of a delayed contribution, referred to as the Raman response. The imaginary part of its Fourier transform, referred to as the Raman gain spectrum, leads to inter-channel stimulated Raman scattering (ISRS). ISRS is a nonlinear effect that redistributes optical power from high to lower frequencies during propagation. However, as the nonlinear fiber response is causal, the Raman spectrum obeys the Kramers-Kronig relations resulting in the real part of the complex valued Raman spectrum. While the impact of the imaginary part (i.e. ISRS) is well studied, the direct implications of its associated real part on the NLI are unexplored. 
\par 
\ 
In this work, a theory is proposed to analytically quantify the impact of the real Raman spectrum on the nonlinear interference power. Starting from a generalized Manakov equation, an extension of the ISRS Gaussian Noise (GN) model is derived to include the real Raman spectrum and, thus, to account for the complete nonlinear Raman response. Accurate integral expressions are derived and approximations in closed-form are proposed. Different formulations for the case of single -and dual polarized signals are derived and novel analytical approximations of the real Raman spectrum are proposed. Moreover, it is analytically shown that the real Raman spectrum scales the strength of the instantaneous nonlinear distortions depending on the frequency separation of the interacting frequencies. A simple functional form is derived to assess the scaling of the NLI strength. The proposed theory is validated by numerical simulations over C-and C+L band, using experimentally measured fiber data. 
\end{abstract}

\begin{IEEEkeywords}
Optical fiber communications, Gaussian noise model, Nonlinear interference, nonlinear distortion, Stimulated Raman Scattering, First-order perturbation, C+L band transmission, Raman spectrum, delayed nonlinear response, Kramers-Kronig relations
\end{IEEEkeywords}

\IEEEpeerreviewmaketitle

\section{Introduction}

\IEEEPARstart{T}{he} optical fiber is a nonlinear transmission medium and as such imposes unique challenges on modern coherent transmission systems. Within the conventional transmission band (C-band, approximately 5~THz wide), the nonlinear response is typically considered to be instantaneous with respect to the optical field. In general, the instantaneous nonlinear response leads to a power dependent phase shift in the time domain, and through the interaction with chromatic dispersion, to nonlinear signal fluctuations in amplitude and phase. In the context of coherent transmission, these nonlinear signal fluctuations are often modelled as an additional source of noise, denoted nonlinear interference (NLI) \cite{Splett_1993_utc,Tang_2002_tcc,Poggiolini_2012_tgm}. Although the NLI can be partially mitigated \cite{Pepper_1980_cfp, Ip_2008_cod, Secondini_2012_afc, Dar_2014_nbo,Hasegawa_1993_ec,Yousefi_2014_itu} and its fundamental limitation on channel capacity is still unkown \cite{Secondini_2017_sal}, the NLI and its estimation is of significant practical relevance for modern optical transmission systems. 
\par 
\ 
As a low complexity alternative to lengthy numerical split-step simulations, numerous analytical models have been proposed to estimate the power (and more) of the NLI and to yield physical insight on underlying parameter dependencies \cite{Splett_1993_utc, Tang_2002_tcc, Poggiolini_2012_tgm, Johannisson_2013_pao ,Chen_2010_cef, Mecozzi_2012_nsl, Secondini_2012_afc,Dar_2013_pon, Serena_2015_ate, Golani_2016_mtb, Ghazisaeidi_2017_ato, Ghazisaeidi_2019_toc, Vannucci_2002_tpm}. Analytical perturbation models offer approximate solutions of the propagation equations and gained substantial popularity in the last two decades. These models exhibit low computational complexity, high reproducibility and enable efficient system design \cite{Hasegawa_2017_ofd}, analytical throughput estimations \cite{Semrau_2016_air,Shevchenko_2016_air,Bosco_2011_aro}, physical layer aware networking \cite{Anagnostopoulos_2007_pli} and the derivation of novel nonlinearity mitigation schemes \cite{Secondini_2012_afc, Dar_2014_nbo}. Within the C-band, where the nonlinear response is often considered instantaneous, extensive numerical and experimental validations were carried out, approving analytical perturbation approaches \cite{Nespola_2014_gvo, Nespola_2015_evo, Galdino_2016_edo, Cai_2014_tpo}. 
\par 
\ 
However, models based on the conventional nonlinear Schr\"odinger (NLSE) or Manakov (ME) equation, that assume an instantaneous nonlinear response, cannot be applied to transmission bandwidths beyond the C-band. For optical bandwidths beyond the C-band, the delayed part of the nonlinear response, denoted the Raman response, becomes significant and cannot be neglected. In fact, the Raman response may also be relevant within the C-band, depending on accuracy requirements and system parameters. The Raman response is a fundamental property of the guiding medium originating from molecular vibrations and is fully described by its spectrum, referred to as the Raman spectrum. The \textit{imaginary} part of the Raman spectrum corresponds to the well-known Raman \textit{gain} spectrum, as originally measured by Stolen and Ippen in 1973 \cite{Stolen_1973_rgi}. The Raman \textit{gain} spectrum characterizes the power transfer between two frequencies as a function of their frequency separation. It leads to inter-channel stimulated Raman scattering (ISRS) which is a nonlinear effect that redistributes optical power from high to lower frequencies during propagation, where the power evolution is described by the Raman gain equations \cite{Tariq_1993_acm}.
\par 
\ 
While it is known how to include the Raman response into numerical split-step simulations since 1989, based on the generalized NLSE (GNLSE) \cite{Blow_1989_tdo}, analytical perturbation models that include the imaginary Raman spectrum were only recently introduced \cite{Semrau_17_ard}. Analytical perturbation models were extended to account for the imaginary Raman spectrum by modelling the effect of ISRS. In particular, as the imaginary Raman spectrum results in ISRS, it can be accounted for by introducing a signal power profile (i.e. a suitable loss function), such that the profile resembles the power transfer caused by ISRS. Mathematically, the signal power profile is obtained from the Raman gain equations \cite{Tariq_1993_acm} or their analytical approximations \cite{Zirngibl_1998_amo, Christodoulides_1996_eos}. The approach neglects the temporal gain dynamics of ISRS which, although not yet analytically proven, seem to be negligible \cite{Forghieri_1995_eom,Ho_2000_spo,Saavedra_2018_isr,Minoguchi_2018_eos}. 
\par 
\ 
Applying that methodology, the conventional Gaussian Noise (GN) model \cite{Splett_1993_utc,Tang_2002_tcc,Poggiolini_2012_tgm} was extended to account for ISRS, denoted ISRS GN model, by either introducing effective attenuation coefficients \cite{Semrau_17_ard, Cantono_2018_mti} or by deriving a GN model subject to a generic signal power profile \cite{Roberts_17_cpo, Semrau_2018_tgn, Semrau_2018_tig, Cantono_2018_mti, Cantono_2018_oti}. In addition to accurate integral formulations, fast closed-form approximations have been proposed to enable real-time performance estimations \cite{Semrau_17_ard,Semrau_2018_aca,Poggiolini_2018_agg,Semrau_2019_aca}. A quantitative comparison between both approaches in integral and closed-form can be found in \cite{Semrau_2019_oac}. Furthermore, the link function of the ISRS GN model can be combined with the formalism derived in \cite{Carena_2014_emo}, to account for arbitrary (non-Gaussian) modulation formats \cite{Semrau_2019_amf,Serena_2020_ons,Lasagni_2019_are, Rabbani_2019_aga}. To avoid any additional computational complexity, as often associated with modulation format aware models, a modulation format correction in closed-form was derived in \cite{Semrau_2019_amf,Semrau_2019_mfd}. Although, some features of NLI (e.g. correlation time, phase noise etc.) and their relations to ISRS are still unexplored, the imaginary part of the complex Raman spectrum has been successfully integrated in the analytical modelling of the nonlinear distortions. 
\par 
\ 
However, as the Raman response is a causal response function, its spectrum obeys the Kramers-Kronig (KK) relations leading to a non-zero real part of the complex Raman spectrum \cite{Hellwarth_1975_oaf,Stolen_1989_rrf,Lin_2006_rrf}. While the impact of the imaginary part is well known (leading to an ISRS power transfer), studies about the real Raman spectrum and its consequences are less prevalent. In \cite{Stolen_1989_rrf,Lin_2006_rrf}, it was shown that the real Raman spectrum changes the nonlinear refractive index, indicating changes on the nonlinear interference. However, to the best of our knowledge, the actual qualitative and quantitative impact of the real Raman spectrum on the NLI is unknown and has not been investigated. 
\par 
\ 
In this work, a theory is proposed to numerically and analytically assess the impact of the Raman response on the NLI, with particular emphasis on the real part of the complex valued Raman spectrum. First, the generalized Manakov equation (GME) is introduced that is suitable to study a delayed nonlinear response to all orders and which serves as a basis for a first-order regular perturbation (RP1) approach. Formulas to estimate the fractional contribution of the Raman response on the total nonlinear response are presented and a novel analytical approximation of the real Raman spectrum is proposed. 
\par 
\ 
The key contributions of this work are twofold: First, the RP1 approach is applied to the GME, extending the ISRS GN model to account for the real Raman spectrum and, thus, accounting for the complete Raman response. Second, enabled by a closed-form approximation of the ISRS GN model, the impact of the real Raman spectrum on the NLI is found analytically. It is shown that the real Raman spectrum scales the NLI depending on the frequency separation of the interacting frequencies. It is also shown that the relative NLI scaling is only a function of the real Raman spectrum itself. Additionally, it is demonstrated that the NLI scaling is different for single and dual polarized signals. 
\par 
\ 
Finally, the proposed model is validated by numerical simulations over the C-band (5~THz) and the C+L band (10~THz). Transmission scenarios where the real Raman spectrum is relevant, and where its impact can be neglected, are identified. 
\newline \newline
The remainder of this paper is organized as follows: In Sec. \ref{sec:GME} the general formalism is introduced to numerically model the impact of a delayed nonlinear response on the propagating electrical field. The fractional contribution of the Raman response is addressed in Sec. \ref{sec:delayed_response} and a novel analytical approximation of the complex Raman spectrum is proposed in Sec. \ref{sec:analytical_approximation}. In Sec. \ref{sec:extending_ISRS_GN_model}, the ISRS GN model is extended to account for the real Raman spectrum. Approximations in closed-form and scaling factors, that analytically assess the impact of the real Raman spectrum, are derived in Sec. \ref{sec:cf_approx}. The results are then used in Sec. \ref{sec:impact_real_Raman} to analytically show the impact of the real Raman spectrum on the NLI, where the special case of single polarization is discussed in Sec. \ref{sec:single_pol}. Finally, the derived theory is numerically validated in Sec. \ref{sec:validation}, with an optical bandwidth of up to 5~THz in Sec. \ref{sec:results_C-band} and of up to 10~THz in Sec. \ref{sec:results_CL-band1} and \ref{sec:results_CL-band2}.
\section{The delayed nonlinear fiber response}
\label{sec:GME}
This section covers the mathematical framework to study the impact of a delayed nonlinear response on the propagating optical wave. In \cite{Blow_1989_tdo} the NLSE was generalized to account for a delayed nonlinear response for single polarization, which the authors refer to as the generalized NLSE (GNLSE). Similar to the GNLSE, the Manakov equation can be generalized to account for a delayed nonlinear response, referred to as the generalized Manakov equation (GME) and written as 
\begin{equation}
\begin{split}
&\frac{\partial}{\partial z}E_\text{x}\left(z,t\right)=\left(-\frac{\alpha}{2}-j\frac{\beta_2}{2}\frac{\partial^2}{\partial t^2}+\frac{\beta_3}{6}\frac{\partial^3}{\partial t^3}\right)E_\text{x}\left(z,t\right) \\
&+j\gamma E_\text{x}\left(z,t\right)\int h\left(\tau\right)\left[\left|E_\text{x}\left(z,t-\tau\right)\right|^2+\left|E_\text{y}\left(z,t-\tau\right)\right|^2\right]d\tau,
\label{eq:GME_time}
\end{split}
\end{equation}
where $E_\text{x}\left(z,t\right)$ and $E_\text{y}\left(z,t\right)$ are the complex envelopes of the electric field in x and y-polarizations, $\alpha$ is the attenuation coefficient, $\beta_2$ is the group velocity dispersion (GVD), $\beta_3$ is the GVD slope, $\gamma$ is the nonlinearity coefficient and $h\left(t\right)$ is the nonlinear impulse response. The nonlinear response consists of an instantaneous part, caused by electronic contributions, and a delayed part, caused by molecular vibrations. The nonlinear impulse response is written as \cite{Agrawal_2012_nfo, Antonelli_2016_mon}
\begin{equation}
\begin{split}
&h(t) =\frac{8}{9} \left(1-f_r\right)\delta\left(t\right) + f_rh_r(t),
\label{eq:response_time}
\end{split}
\end{equation}
with the fractional contribution of the Raman response $f_r$ and the Raman response itself $h_r(t)$. The Fourier transform of the Raman response $h_r(t)$ is referred to as the complex valued Raman spectrum $H_r(f)=\mathcal{F}\left\{h_r(t)\right\}$. The fractional contribution of the Raman (delayed) response $f_r$ is addressed in more detail in Sec. \ref{sec:delayed_response}.
\par 
\ 
The GME \eqref{eq:GME_time} and the nonlinear impulse response written as in \eqref{eq:response_time} extend the result in \cite[Eq. (103)]{Antonelli_2016_mon} to account for all orders of the delayed optical field and is similar to the GNLSE in the case of single polarization \cite{Blow_1989_tdo}. In \cite{Antonelli_2016_mon}, it was shown that, for dual polarized signals, rapidly varying birefringences only scale the instantaneous part of the nonlinear fiber response by the well-known factor $\frac{8}{9}$ as written in Eq. \eqref{eq:response_time}. In Appendix \ref{apx:sec:GME_and_antonelli}, it is shown that the GME reduces to the result in \cite[Eq. (103)]{Antonelli_2016_mon}, when the delayed optical field is approximated to first-order, i.e. $E\left(t-\tau\right)= E\left(t\right)-\tau\frac{\partial}{\partial t}E\left(t\right)$. It is also shown that this approximation is equivalent to a zeroth-order approximation of the real and a first-order approximation of the imaginary part of the nonlinear response. This corresponds to a constant real, and linear imaginary Raman spectrum. However, to study the impact of the functional shape of the real Raman spectrum, higher orders must be included and Eqs. \eqref{eq:GME_time}\eqref{eq:response_time} serve as a basis for the remainder of this work. 
\par 
\ 
It is useful to analyse the GME \eqref{eq:GME_time} in the frequency domain\footnote{In this work, the Fourier transform is defined as in \cite{Agrawal_2012_nfo}, i.e., the forward transform is given by $\mathcal{F}\left\{f\left(t\right)\right\}\left(f\right)=\int f\left(t\right)e^{j2\pi f t}dt $.} due to the convolution in the nonlinear term. The GME in the frequency domain is 
\begin{equation}
\begin{split}
&\frac{\partial}{\partial z}E_\text{x}\left(f\right)=\widetilde{\Gamma}(f)E_\text{x}\left(f\right) \\
&+j\gamma E_\text{x}\left(f\right)*H(f)\left[E_\text{x}(f)*E_\text{x}^*(-f)+E_\text{y}(f)*E_\text{y}^*(-f)\right]
\label{eq:GME_freq}
\end{split}
\end{equation}
with complex propagation constant $\widetilde{\Gamma}(z,f)=\frac{g\left(z,f\right)}{2}+j2\pi^2\beta_2f^2+j\frac{4}{3}\pi^3\beta_3f^3$ and $g\left(z,f\right)$ is an arbitrary loss profile which can be used to analytically model the impact of inter-channel stimulated Raman scattering (ISRS) and which arises from the imaginary part of the Raman response. To obtain \eqref{eq:GME_time}, the loss profile is $g\left(z,f\right)=-\alpha$. The nonlinear transfer function is 
\begin{equation}
\begin{split}
&H(f) =\underbrace{\frac{8}{9} \left(1-f_r\right)}_\text{instantaneous} + \underbrace{f_rH_r(f)}_\text{delayed},
\label{eq:response_freq}
\end{split}
\end{equation}
with the complex valued Raman spectrum as
\begin{equation}
\begin{split}
H_r\left(f\right)&=\underbrace{n_r\left(f\right)}_{\text{real part}} + j\underbrace{g_r\left(f\right)}_{\text{ISRS}},
\label{eq:delayed_response_frequency}
\end{split}
\end{equation}
where $n_r\left(f\right)$ and $g_r\left(f\right)$ are the real and imaginary parts of the Raman spectrum. The imaginary part of the Raman response leads to inter-channel stimulated Raman scattering which amplifies low frequencies at the expense of high frequency components during propagation. The real part of the Raman spectrum originates from the causality principle of the nonlinear response \cite{Hellwarth_1975_oaf,Stolen_1989_rrf,Lin_2006_rrf}. The imaginary Raman spectrum (the Raman \textit{gain} spectrum) can be obtained from fiber measurements. The complex Raman spectrum as in \eqref{eq:delayed_response_frequency} is connected to the measured Raman \textit{gain} spectrum as  
\begin{equation}
\begin{split}
n_r\left(f\right) &= \frac{1}{2A_\text{eff}f_r\gamma}\tilde{n}_r\left(f\right)=\frac{\lambda_0}{4\pi f_r n_2}\tilde{n}_r\left(f\right)\\
g_r\left(f\right) &= \frac{1}{2A_\text{eff}f_r\gamma} \tilde{g}_r\left(f\right)=\frac{\lambda_0}{4\pi f_r n_2}\tilde{g}_r\left(f\right)
\label{eq:spec_prefactors}
\end{split}
\end{equation}
where $A_\text{eff}$ is the effective core area, $\lambda_0$ is the reference wavelength, $n_2$ is the nonlinear refractive index and $\tilde{g}_r\left(f\right)$ is the co-polarized Raman gain spectrum, that is obtained from measurements. To show that \eqref{eq:spec_prefactors} is the correct normalization ($H_r\left(f\right)$ is dimensionless), the Raman gain equations are derived from the GME in Appendix \ref{sec:apx:GMEtoODE}.
\par 
\ 
It is sufficient to measure the real \textit{or} the imaginary part of the Raman spectrum as both are related through the Kramers-Kronig (KK) relations, originating from the causality principle. Using the KK relations, the real part of the co-polarized Raman gain spectrum is 
\begin{equation}
\begin{split}
\tilde{n}_r\left(f\right) = \mathcal{H}\left\{\tilde{g}_r\left(f\right)\right\} = \frac{1}{\pi}\text{p.v.}\int \frac{\tilde{g}_r\left(f'\right)}{f'-f}df'
\label{eq:kk_relations}
\end{split}
\end{equation}
where $\mathcal{H}\left\{\cdot\right\}$ denotes the Hilbert transform and $\text{p.v.}$ denotes the Cauchy principal value. The functional shape of the complex valued Raman spectrum is further addressed in Sec. \ref{sec:analytical_approximation}.
\subsection{The fractional contribution of the Raman response $f_r$}
\label{sec:delayed_response}
In this section the fractionional contribution of the Raman response $f_r$ is discussed in more detail. The quantity $f_r$ can be obtained directly from the real Raman spectrum. This is because the Raman response $h_r\left(t\right)$ has a fractional contribution on the instantaneous part of the nonlinear response. To examine this for single polarization, we analyse Eq. \eqref{appendixA:eq:GME_taylor} (with  $E_\text{y}\left(z,t\right)=0$, $n=0$ and $\frac{8}{9}\to 1$) which is the GNLSE only considering instantaneous effects on the electrical field, mathematically $E\left(z,t-\tau\right)=E\left(z,t\right)$ and resulting in
\begin{equation}
\begin{split}
&\frac{\partial}{\partial z}E_\text{x}\left(z,t\right)=\left(-\frac{\alpha}{2}-j\frac{\beta_2}{2}\frac{\partial^2}{\partial t^2}+\frac{\beta_3}{6}\frac{\partial^3}{\partial t^3}\right)E_\text{x}\left(z,t\right) \\
&+j\gamma E_\text{x}\left(z,t\right)\left|E_\text{x}\left(z,t\right)\right|^2\left[\left(1-f_r\right) + \frac{1}{2A_\text{eff}\gamma}\tilde{n}_r\left(0\right)\right].
\label{eq:fr_normalisation}
\end{split}
\end{equation}
Eq. \eqref{eq:fr_normalisation} shows that, while the imaginary part of the Raman response vanishes, the real part imposes a contribution on the instantaneous nonlinear response  with a constant value $\tilde{n}_r\left(0\right)$. This means that the real part, evaluated at $f=0$, is present for any spectral distribution of the electrical field and that it cannot be distinguished from quasi-instantaneous contributions of the nonlinear response. In other words, $\tilde{n}_r\left(0\right)$ directly contributes to the nonlinear refractive index $n_2$ (and in turn to the nonlinearity coefficient) \cite{Stolen_1989_rrf}. As a consequence, experimental measurements of the nonlinearity coefficient (or the nonlinear refractive index) measure the instantaneous contribution of both, the instantaneous and the Raman response. This motivates the normalization of the Raman response such that 
\begin{equation}
\begin{split}
&1\overset{!}{=}\left(1-f_r\right) + \frac{1}{2A_\text{eff}\gamma}\tilde{n}_r\left(0\right),
\label{eq:fr_normalisation_condition}
\end{split}
\end{equation}
and we obtain the fractional contribution of the Raman response
\begin{equation}
\begin{split}
&f_r=\frac{\lambda_0}{4\pi n_2}\tilde{n}_r\left(0\right).
\label{eq:fr}
\end{split}
\end{equation}
Additionally, Eq. \eqref{eq:fr} yields a normalization of the real Raman spectrum such that $n_r\left(0\right)=1$. 
\par 
\ 
The complex Raman spectrum of a Corning\textsuperscript{\textregistered} SMF-28\textsuperscript{\textregistered} ultra low loss (ULL) fiber is shown in Fig. \ref{fig:raman_spectrum}. Assuming $\lambda_0=1550$~nm and $n_2=2.1\cdot 10^{-20}$~$\text{m}^2\text{W}^{-1}$ (a typical value for silica-core fibres \cite{Makovejs_2016_tst}), the fractional contribution of the Raman response is $f_r=0.23$. This means that around 23\% of the nonlinearity coefficient $\gamma$ stems from the Raman response and 77\% stems from the quasi-instantaneous, electronic contributions. A value of $n_2=2.6\cdot 10^{-20}$~$\text{m}^2\text{W}^{-1}$ yields a fractional contribution of $f_r=0.18$, which is consistent with the values reported in \cite{Agrawal_2012_nfo,Stolen_1989_rrf}
\par 
\ 
The normalisation of the Raman response was carried out with respect to the single polarization case. If the polarisation state changes randomly along the fiber length, as described by the GME \eqref{eq:GME_time}, only the instantaneous part of the nonlinear response is reduced by a factor of $\frac{8}{9}$ in \eqref{eq:response_time}. This results in a different normalization of the Raman response with respect to the GME, or in a sightly different measurement of $n_2$ in long fiber lengths \cite{Antonelli_2016_mon}. However for the remainder of this manuscript, we assume that $n_2$ was evaluated according to the single polarization case, e.g. by measuring the nonlinear refractive index for short or polarization-maintaining fibers and therefore corresponds to a normalization as in \eqref{eq:fr}. Only accounting for instantaneous nonlinear effects on the electrical field in the GME leads to 
\begin{equation}
\begin{split}
&\frac{\partial}{\partial z}E_\text{x}\left(z,t\right)=\left(-\frac{\alpha}{2}-j\frac{\beta_2}{2}\frac{\partial^2}{\partial t^2}+\frac{\beta_3}{6}\frac{\partial^3}{\partial t^3}\right)E_\text{x}\left(z,t\right) \\
&+j\frac{8+f_r}{9}\gamma E_\text{x}\left(z,t\right)\left[\left|E_\text{x}\left(z,t\right)\right|^2+\left|E_\text{y}\left(z,t\right)\right|^2\right],
\end{split}
\end{equation}
which reduces to the conventional Manakov equation if the Raman response is neglected, i.e. $f_r=0$ \cite{Antonelli_2016_mon}. It should be added that one could introduce an effective nonlinearity coefficient as $\gamma_\text{eff}=\frac{8+f_r}{8}\gamma$ for dual polarization signals, to resemble the conventional Manakov equation and in case $\gamma_\text{eff}$ was obtained from dual polarization measurements in long fibers. 
\subsection{Analytical approximation of the complex Raman spectrum $H_r\left(f\right)$}
\label{sec:analytical_approximation}
\begin{figure}[h]
   \centering
    \includegraphics[]{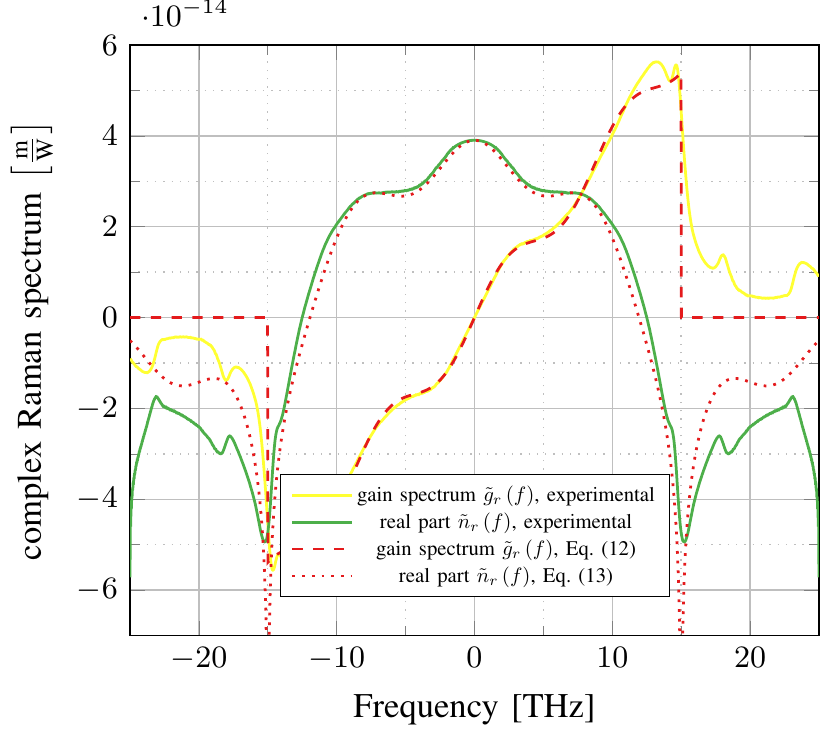}
\caption{\small The real and imaginary part of the complex Raman spectrum $\tilde{H}_r\left(f\right)=\tilde{n}_r\left(f\right) + j\tilde{g}_r\left(f\right)$ of a SMF-28 ULL fiber. The spectrum was obtained from experimental measurements of the imaginary part and using the proposed analytical approximations Eqs. (12) and (13).}
\label{fig:raman_spectrum}
\end{figure}
In this section, a novel analytical approximation for the complex Raman spectrum is proposed. Due to the one-to-one relationship of the KK relations, a suitable analytical approximation can be found for either the real or the imaginary part. A very effective analytical description of the imaginary part (i.e. the Raman \textit{gain} spectrum) is given by the triangular approximation, which is a linear regression of the imaginary part and valid up to 15~THz \cite{Chraplyvy_1984_opl}. While significantly more complex (and more accurate) analytical approximations exist, only the triangular approximation leads to an analytical solution of the Raman gain equations to obtain a closed-forn formula for the ISRS power transfer \cite{Christodoulides_1996_eos, Zirngibl_1998_amo}. This solution can then be used to obtain analytical formulas in integral and closed-form for estimating the NLI in the presence of inter-channel stimulated Raman scattering \cite{Semrau_2018_tgn, Semrau_2018_tig, Semrau_2019_aca}. 
\par 
\ 
The functional shape of the imaginary Raman spectrum arises from the amorphous structure of silica and can be separated into an isotropic (direction independent) and a less significant anisotropic (direction dependent) contribution, where the isotropic part is responsible for the entire triangular-like shape and the anisotropic part is solely responsible for the local peak at around 3~THz, known as the Boson peak \cite{Lin_2006_rrf}. The isotropic part, that is responsible for the majority of Raman gain spectrum, is well modelled by the triangular approximation or other analytical approximations based on damped-harmonic oscillators and Lorentzian profiles \cite{Stolen_1989_rrf, Hook_1994_sia, Blow_1989_tdo}. While both approaches are suitable for modelling the imaginary part, they do not sufficiently describe the Boson peak around 3~THz, which is particularly important for the real Raman spectrum. 
\par 
\ 
For the imaginary part (i.e. for ISRS) an accurate modelling for large frequency separations ($>5$~THz) is important due to the large Raman gain in that region. However for the real part, good accuracy for low frequency separations is required (which will be clearer in Sec. \ref{sec:impact_real_Raman}) where the Boson peak has a pronounced impact. As a result, the triangular and single Lorentzian approximations of the imaginary part are not suitable to analytically approximate the real Raman spectrum. 
\par 
\ 
In \cite{Lin_2006_rrf} an approximation for the Boson peak was proposed using a single functional form in the time domain. In \cite{Hollenberg_2002_mvm}, 13 vibrational modes were introduced to accurately fit the entire complex Raman response in the time domain. Although both approaches are suitable for modelling the real part, we adopt a different fitting approach in this work. Similar to the triangular approximation, we seek a simple representation directly in the frequency domain that is sufficiently accurate for the use in analytical perturbation models. For this purpose, we superimpose the triangular approximation with a suitable sine wave, mathematically 
\begin{equation}
\begin{split}
&\tilde{g}_r\left(f\right)=\left[\tilde{C}_r\cdot f  + \tilde{A}_r\cdot\text{sin}\left(\tilde{\omega}_r f\right)\right]\Pi \left(\frac{f}{\tilde{B}_r}\right),
\label{eq:raman_gain_spectrum_analytical}
\end{split}
\end{equation}
with $\Pi \left(x\right)$ denoting the rectangular function and $\tilde{B}_r=30$~THz. The sine wave in \eqref{eq:raman_gain_spectrum_analytical} leads to a more accurate description of the imaginary part for low frequency separations, necessary to account for the Boson peak. Using the KK relations \eqref{eq:kk_relations}, the imaginary part \eqref{eq:raman_gain_spectrum_analytical} can be transformed to its corresponding real part. This is done in Appendix \ref{sec:analytical_real_part} with the result as 
\begin{equation}
\begin{split}
\tilde{n}_r\left(f\right)&=\frac{\tilde{C}_r}{\pi}\left[ f\log\left(\left|\frac{2f-\tilde{B}_r}{2f+\tilde{B}_r}\right|\right)+\tilde{B}_r \right]\\
&+ \tilde{A}_r\cdot\text{cos}\left(\tilde{\omega}_r f\right)+\tilde{D}_r.
\label{eq:real_part_analytical}
\end{split}
\end{equation}
The parameter $\tilde{D}_r$ was introduced to offset the mismatch of the integration domain between \eqref{eq:raman_gain_spectrum_analytical} and the actual Raman gain function, which leads to a vertical offset. While the parameters in \eqref{eq:real_part_analytical} have physical origins and their values could potentially be derived, we choose to treat them as fitting parameters. In particular, the parameters are fitted using the impact of the real part on the XPM contribution of the NLI as cost function (addressed in Sec. \ref{sec:impact_real_Raman}). 
\par 
\
Fig. \ref{fig:raman_spectrum} shows the real and imaginary part of the Raman spectrum obtained from measurements of a SMF-28 ULL and their analytical approximations given by \eqref{eq:real_part_analytical} and \eqref{eq:raman_gain_spectrum_analytical}. The fitting parameters are $\tilde{C}_r=3.87\cdot 10^{-27}$~$\frac{\text{m}}{\text{W}\text{Hz}}$. $\tilde{B}_r=30\cdot 10^{12}$~Hz, $\tilde{A}_r=4.2\cdot 10^{-15}$~$\frac{\text{m}}{\text{W}}$, $\tilde{\omega}_r=7.20\cdot 10^{-13}$~$\text{rad}\cdot \text{Hz}$ $\tilde{D}_r=-2.12\cdot 10^{-15}$~$\frac{\text{m}}{\text{W}}$.  The analytical approximation proposed in this section is in excellent agreement with the experimental measurements. This is particularly true for the Boson peak in the real part around $f=0$, making \eqref{eq:real_part_analytical} ideal for modelling the impact of the real Raman spectrum on the NLI.
In Sec \ref{sec:impact_real_Raman}, the goodness of fit of \eqref{eq:real_part_analytical} is evaluated with respect to NLI predictions. 
\section{Extending the ISRS GN model}
\label{sec:extending_ISRS_GN_model}
In this section, the ISRS GN model is extended to analytically account for the complete Raman response. As discussed in Sec. \ref{sec:GME}, the imaginary Raman spectrum leads to inter-channel stimulated Raman scattering, redistributing power across the optical spectrum during propagation. In our previous work, we derived the ISRS GN model that is capable to account for ISRS and for the imaginary Raman spectrum. We derived accurate integral formulations \cite{Semrau_17_ard,Semrau_2018_tgn, Semrau_2018_tig} and a closed-form approximation for Gaussian constellations \cite{Semrau_2019_aca} and arbitrary modulation formats \cite{Semrau_2019_amf, Semrau_2019_mfd}. In the following, these results are extended to account for the real Raman spectrum and, thus, for the complete Raman response. 
\par 
\ 
In the following, the general formalism of analytical performance estimation is briefly introduced. After coherent detection and electronic dispersion compensation, the channel dependent signal-to-noise ratio (SNR) can be calculated as
\begin{equation}
\begin{split}
\label{eq:SNR}
\textnormal{SNR}_i \approx \frac{P_i}{nP_\textnormal{ASE} + \eta_n P_i^3},
\end{split}
\end{equation}
where $P_i$ is the launch power of channel $i$, $P_\textnormal{ASE}$ is the amplified spontaneous emission (ASE) noise power over the channel bandwidth and $\eta_n$ is the nonlinear interference coefficient after $n$ spans. When the channel bandwidth $B_{\text{ch}}$ is small compared to the total optical bandwidth $B_\text{tot}$, the power spectral density (PSD) of the NLI can be considered locally flat and $\eta_n$ can be approximated as
\begin{equation}
\begin{split}
\label{eq:eta_conversation}
\eta_n\left(f_i\right) = \int_{-\frac{B_{\text{ch}}}{2}}^{\frac{B_{\text{ch}}}{2}}\frac{G\left(\nu+f_i\right)}{P_i^{3}}d\nu \approx \frac{B_{\text{ch}}}{P_i^{3}}G\left(f_i\right),
\end{split}
\end{equation}
where $f_i$ is the center frequency of channel $i$. The PSD of the nonlinear interference $G\left(f\right)$ is obtained from analytical perturbation approaches.
\par 
\ 
The extension of the ISRS GN model, to account for the real Raman spectrum, for dual polarization is derived in Appendix \ref{sec:apx:derivation}, with the result as 
\begin{equation}
\begin{split}
&G(f) =\frac{16}{27}\gamma^2 \int \int  df_1df_2 \ G_{\text{Tx}}(f_1)G_{\text{Tx}}(f_2)G_{\text{Tx}}(f_1+f_2-f)\\
&\cdot\left|\int_0^Ld\zeta\ \sqrt{\frac{\rho(\zeta,f_1)\rho(\zeta,f_2)\rho(\zeta,f_1+f_2-f)}{\rho(\zeta,f)}}e^{j\phi\left(f_1,f_2,f,\zeta\right)}\right|^2
\\
&\cdot \left[2R^2\left(f-f_1\right) +R\left( f-f_1\right)R\left(  f-f_2\right)\right],
\label{eq:ISRSGNmodel_rho}
\end{split}
\end{equation}
with 
\begin{equation}
\begin{split}
&R(f) =\frac{9}{8\sqrt{3}}\Re \left\{H(f)\right\},
\label{eq:Rf}
\end{split}
\end{equation}
and $\phi=-4\pi^2\beta_2\left[(f_1-f)(f_2-f)+\pi\beta_3(f_1+f_2)\right]\zeta$. The variable $\rho(z,f)$ is the normalized signal power profile (e.g.  $\rho(z,f)=e^{-\alpha z}$ in the absence of ISRS), which accounts for the imaginary Raman spectrum and ISRS. The normalized power profile can be obtained from the Raman gain equations or directly from \eqref{eq:GME_time}, as shown in Appendix \ref{sec:apx:GMEtoODE}. For more details on the signal power profile and the ISRS term, the reader is referred to \cite[Sec. II]{Semrau_2018_tgn}. Using the triangular approximation of the imaginary Raman spectrum, as in \eqref{eq:raman_gain_spectrum_analytical} with $A_r=0$, yields an analytical form of the ISRS GN model as 
\begin{equation}
\begin{split}
&G(f) =\frac{16}{27}\gamma^2 \int \int  df_1df_2 \ G_{\text{Tx}}(f_1)G_{\text{Tx}}(f_2)G_{\text{Tx}}(f_1+f_2-f)\\
&\cdot \left| \int_0^L d\zeta \ \frac{P_{\text{tot}}e^{-\alpha \zeta-P_{\text{tot}}C_{\text{r}} L_{\text{eff}}(f_1+f_2-f)}}{\int G_{\text{Tx}}(\nu)e^{-P_{\text{tot}}C_{\text{r}} L_{\text{eff}}\nu} d\nu}e^{j\phi\left(f_1,f_2,f,\zeta\right)}\right|^2\\
&\cdot \left[2R^2\left(f-f_1\right) + R\left( f-f_1\right)R\left(  f-f_2\right)\right],
\label{eq:ISRSGNmodel_Cr}
\end{split}
\end{equation}
where $C_r =\frac{1}{2A_\text{eff}} \tilde{C}_r$ is the regression slope of the polarization averaged Raman gain profile normalized by the effective core area. Both, \eqref{eq:ISRSGNmodel_rho} and \eqref{eq:ISRSGNmodel_Cr} are derived for a single span that can be trivially extended for multi-span systems using the phased array term \cite[Eq. (5)]{Semrau_2018_tgn} or the result in \cite{Semrau_2018_tig} for non-repetitive spans. 
\par 
\ 
The novel contribution in Eqs. \eqref{eq:ISRSGNmodel_rho}\eqref{eq:ISRSGNmodel_Cr} are the terms that involve $R\left(f\right)$ which accounts for the real Raman spectrum. The normalization factor in \eqref{eq:Rf} was introduced for convenience, such that Eqs. \eqref{eq:ISRSGNmodel_rho}\eqref{eq:ISRSGNmodel_Cr} have the same prefactor as standard GN model approaches for dual polarization (i.e. $3\frac{2}{2^3}\frac{8^2}{9^2}=\frac{16}{27}$). It should be noted that $\Re$ $\left\{H(f)\right\}$ contains the real part of the instantaneous and the Raman response as in \eqref{eq:response_freq}. The Raman spectrum $H_r\left(f\right)$ can be obtained from measurements or from the proposed approximation \eqref{eq:real_part_analytical} to yield a fully analytical model. 
\par 
\ 
In the case of $f_r=0$, Eqs. \eqref{eq:ISRSGNmodel_rho}\eqref{eq:ISRSGNmodel_Cr} are identical to the ISRS GN model proposed in \cite{Semrau_2018_tgn}. For completeness, it should be mentioned that $f_r=0$ physically implies the absence of ISRS which, in the model, has to be set manually by setting $\rho(z,f)=e^{-\alpha z}$ in \eqref{eq:ISRSGNmodel_rho} and $C_r=0$ in \eqref{eq:ISRSGNmodel_Cr}. This is because the imaginary Raman spectrum is modelled via a signal power profile. In the absence of the Raman spectrum and ISRS, the results converge to the conventional GN model \cite{Tang_2002_tcc, Poggiolini_2012_tgm}. 
\section{Approximation in closed-form}
\label{sec:cf_approx}
In this section, the impact of the real Raman spectrum is analytically evaluated using the newly derived model \eqref{eq:ISRSGNmodel_rho}. In particular, scaling factors are derived for the SPM and XPM contribution of the total NLI, such that known closed-form approximations of the ISRS GN model \cite{Semrau_2019_aca,Semrau_2019_amf, Semrau_2019_mfd} can be extended to account for the real Raman spectrum. 
\par 
\ 
It is often useful to analytically extract the SPM and XPM contribution from the total NLI. This approach is referred to as the XPM assumption. The SPM contribution $\eta_\text{SPM}\left(f_i\right)$ is the part of the total NLI that the channel of interest (COI) imposes on itself. The XPM contribution $\eta_\text{XPM}^{\left(k\right)}(f_i)$, on the other hand, is the NLI part that a single interfering channel $k$ located at $f_k$ imposes on the COI located at $f_i$. The entire XPM contribution is then obtained by summing over all interfering channels $\left\{k\in\mathbb{N} \ | \ 1 \leq k\leq N_\text{ch} \ \text{and} \  k\neq i \right\}$. NLI contributions that are \textit{jointly} generated by multiple channels are neglected for this analysis. For more details about the XPM assumption, the reader is referred to \cite[Sec. II.b)]{Semrau_2019_aca}. 
\par 
\ 
The benefit of the XPM assumption is a vast reduction of the integration domain in \eqref{eq:ISRSGNmodel_rho} where the integration carried out over the frequency triplet $\left(f_1,f_2,f_i\right)$, denoting the nonlinear perturbation on frequency component $f_i$ caused by frequencies $f_1$, $f_2$ and $f_i$ itself. For the XPM assumption, it is assumed that the Raman spectrum can be considered invariant over one channel bandwidth (cf. Fig. \ref{fig:raman_spectrum}), mathematically $R\left(f\right) \approx R\left(f+\frac{B_i}{2}\right)$. As are result, the terms involving $R\left(f\right)$ can be taken out of the integration in \eqref{eq:ISRSGNmodel_rho}. The precise integration domain approximations, \textit{only for the terms involving} $R\left(f\right)$, are given in the following: \newline 
For the SPM contribution, the frequency triplet $\left(f_1,f_2,f_i\right)$ can be approximated by 
\begin{equation}
\begin{split}
&\left(f_1+f_i,f_2+f_i,f_i\right) \approx \left(f_i,f_i,f_i\right),\\
&\text{for}  \ f_1\in\left[-\frac{B_i}{2},\frac{B_i}{2}\right]  \ \text{and} \  f_2\in\left[-\frac{B_i}{2},\frac{B_i}{2}\right].
\label{eq:spm_domain}
\end{split}
\end{equation}
For the XPM contribution, two summands arise, where their frequency triplets $\left(f_1,f_2,f_i\right)$ can be approximated by 
\begin{equation}
\begin{split}
\text{i):} \ &\left(f_1+f_i,f_2+f_k,f_i\right) \approx \left(f_i,f_k,f_i\right),\\
&\text{for}  \ f_1\in\left[-\frac{B_i}{2},\frac{B_i}{2}\right] \ \text{and} \  f_2\in\left[-\frac{B_k}{2},\frac{B_k}{2}\right], \\
\text{ii):} \ &\left(f_1+f_k,f_2+f_i,f_i\right) \approx \left(f_k,f_i,f_i\right),\\
&\text{for}  \ f_1\in\left[-\frac{B_k}{2},\frac{B_k}{2}\right] \ \text{and} \  f_2\in\left[-\frac{B_i}{2},\frac{B_i}{2}\right].
\label{eq:xpm_domain}
\end{split}
\end{equation}
Eqs. \eqref{eq:spm_domain}\eqref{eq:xpm_domain} can be used to introduce scaling factors for the SPM and the XPM contribution that entirely capture the effect of the real Raman spectrum. 
\par 
\ 
Under the XPM assumption, the nonlinear interference coefficient after one span can be written as 
\begin{equation}
\begin{split}
&\eta_1\left(f_i\right)\approx R_\text{SPM}\eta_\text{SPM}\left(f_i\right)+\eta_\text{XPM}\left(f_i\right),
\end{split}
\end{equation}
with the total XPM contribution as
\begin{equation}
\begin{split}
&\eta_\text{XPM}\left(f_i\right)=\sum_{k=1,k\neq i}^{N_\text{ch}} R_\text{XPM}\left(\Delta f\right)\eta_\text{XPM}^{\left(k\right)}(f_i).\\
\label{eq:XPM_eta}
\end{split}
\end{equation}
The scaling factors for SPM $R_\text{SPM}$ and XPM $R_\text{XPM}\left(\Delta f\right)$ account for the real Raman spectrum and both are normalized such that $R_\text{SPM}=R_\text{XPM}\left(\Delta f\right)=1$ for $f_r=0$. As a result, the NLI contributions for SPM $\eta_\text{SPM}\left(f_i\right)$ and XPM $\eta_\text{XPM}^{\left(k\right)}(f_i)$ can be directly obtained from known closed-form approximations (e.g. \cite[Eqs. (10-11)]{Semrau_2019_aca}). Utilising Eqs. \eqref{eq:spm_domain}\eqref{eq:xpm_domain}, the scaling coefficients are given by 
\begin{equation}
\begin{split}
&R_\text{SPM}=3R^2\left(0\right),
\label{eq:sclaing_SPM}
\end{split}
\end{equation}
and 
\begin{equation}
\begin{split}
&R_\text{XPM}\left(\Delta f\right)=R^2\left(\Delta f\right)+R\left(\Delta f\right)R\left( 0\right)+R^2\left(0\right),
\label{eq:sclaing_XPM}
\end{split}
\end{equation}
where $\Delta f=\left|f_i-f_k\right|$. Eqs. \eqref{eq:sclaing_SPM}\eqref{eq:sclaing_XPM} combined with the analytical approximation of the real Raman spectrum \eqref{eq:real_part_analytical} lead to a fully analytical model in closed-form to account for the real Raman spectrum, without the need of any numerical computation. 
\section{The impact of the real Raman spectrum on the nonlinear interference}
\label{sec:impact_real_Raman}
\begin{figure}[h]
   \centering
    \includegraphics[]{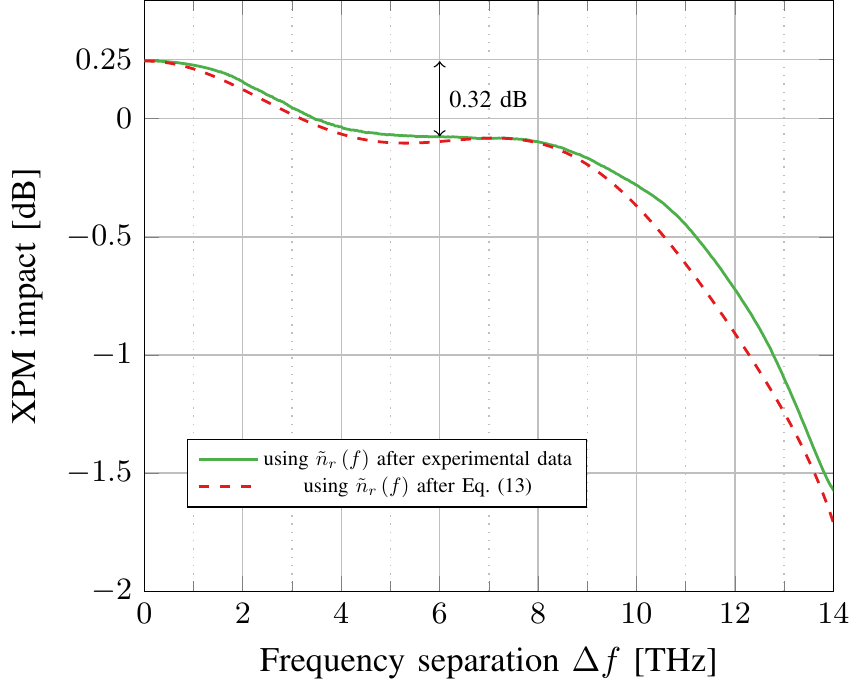}
\caption{\small The impact of the real Raman spectrum ($f_r= 0.23$) with respect to an instantaneous nonlinear response ($f_r=0$) on the XPM contribution of the NLI, after \eqref{eq:XPM_impact}. Note that a freqeuency separation of $\Delta f =0$ represents the SPM contribution of the NLI. Shown are the results obtained from a measured and an analytical approximation of the complex Raman spectrum as shown in Fig. \ref{fig:raman_spectrum}. The parameters of the analytical approximation are listed in Sec. \ref{sec:analytical_approximation}. The results only consider the real part of the complex Raman spectrum and neglect ISRS.}
\label{fig:XPM_impact}
\end{figure}

In this section, the scaling factors, derived in the previous section, are used to assess the impact of the real Raman spectrum on the nonlinear interference power. The imaginary part of the Raman response is modelled via ISRS and the signal power profile $\rho\left(z,f\right)$. Therefore, we define the SPM/XPM impact of the real Raman spectrum with respect to an instantaneous nonlinear response as 
\begin{equation}
\begin{split}
\frac{\eta_\text{XPM}^{\left(k\right)}\left(f_i\right)\Big|_{f_r \neq 0}}{\eta_\text{XPM}^{\left(k\right)}\left(f_i\right)\Big|_{f_r = 0}} &= R_\text{XPM}\left(\Delta f\right).
\label{eq:XPM_impact}
\end{split}
\end{equation}
Remarkably, the only assumption in deriving \eqref{eq:XPM_impact} is that the complex Raman spectrum is constant over a single channel bandwidth. Additionally, \eqref{eq:XPM_impact} converges to the SPM impact for $\Delta f \to 0$. It is therefore very accurate in evaluating the impact of the real Raman spectrum on the SPM and XPM contribution which, in most cases, represent the total amount of the NLI. 
\par 
\ 
The impact of the real Raman spectrum on the NLI, after \eqref{eq:XPM_impact}, as a function of the frequency spacing between interfering channel and channel of interest is shown in Fig. \ref{fig:XPM_impact}. The real Raman spectrum was taken from experimental measurements and from its analytical approximation \eqref{eq:real_part_analytical}, as shown in Fig. \ref{fig:raman_spectrum}. 
\par 
\ 
For a frequency separation of $\Delta f=0$, the real Raman spectrum leads to an increase in NLI. This is a consequence of the fractional Raman contribution $f_r$ and $\tilde{n}_r\left(0\right)$, as discussed in Sec. \ref{sec:delayed_response} and in \cite{Antonelli_2016_mon}. This results in a constant factor (an offset in dB scale) of precisely $R_ \text{XPM}\left(0\right)=3R^2\left(0\right)=\frac{\gamma_\text{eff}^2}{\gamma^2}=\frac{\left(8+f_r\right)^2}{8^2}=0.2456$~dB. However, one of the key contributions of this work is the determination of the impact of the real Raman spectrum for $\Delta f \neq 0$ as the result in \cite[Eq. (103)]{Antonelli_2016_mon} only accounts for $\Delta f = 0$ (i.e. to zeroth-order). 
\par 
\
Fig. \ref{fig:XPM_impact} shows that the real Raman spectrum increases the NLI for closely spaced channels and then decreases the XPM contribution with increasing frequency separation. The reader is reminded that this is \textit{not} due to channel walk-off or dispersive effects, and solely a property of the real Raman spectrum. The NLI is reduced by around 0.32~dB when the frequency separation between COI and INT is 6~THz, and further decreased for increasing frequency separations. The total NLI is (approximately) a summation over all XPM contributions and the impact of the real Raman response is then a weighted sum of the function $R_\text{XPM}\left(\Delta f\right)$.
\par 
\ 
Fig. \ref{fig:XPM_impact} shows a good agreement between the measured real Raman spectrum and its analytical approximation \eqref{eq:real_part_analytical}. In this work, the fitting parameter in \eqref{eq:real_part_analytical} have been chosen to minimize the squared errors of $R_\text{XPM}\left(\Delta f\right)$ for $\Delta f \in \left[0,10\right]$~THz. This has been done as we are interested in a good fit of \eqref{eq:real_part_analytical} with respect to the NLI, rather than in terms of Raman spectrum itself. As a result, \eqref{eq:real_part_analytical} yields an excellent analytical approximation of the real Raman spectrum for NLI predictions. 
\par 
\ 
In conclusion, we analytically quantified that the real Raman spectrum \textit{reduces} the NLI as a function of increasing frequency separation between COI and INT. Additionally, we proposed suitable analytical approximations of the impact of the real Raman spectrum to avoid the need of extensive measurements, enabling cross validations and analytical performance estimations in closed-form.  
\section{Single polarisation case}
\label{sec:single_pol}
In this section, the case of single polarization (e.g. using polarization maintaining fibers) is discussed. For the single polarization case, the evolution of the complex envelope of the electrical field is govered by the generalized nonlinear Schr\"odinger equation \cite{Blow_1989_tdo}. In the context of analytically modelling the real Raman spectrum, the single polarization case requires a separate discussion as its functional form is different. The detailed derivation is carried out in Appendix \ref{sec:apx:derivation} with the result being 
\begin{equation}
\begin{split}
&G(f) =2\gamma^2 \int \int  df_1df_2 \ G_{\text{Tx}}(f_1)G_{\text{Tx}}(f_2)G_{\text{Tx}}(f_1+f_2-f)\\
&\cdot \left| \int_0^L d\zeta \ \frac{P_{\text{tot}}e^{-\alpha \zeta-P_{\text{tot}}C_{\text{r}} L_{\text{eff}}(f_1+f_2-f)}}{\int G_{\text{Tx}}(\nu)e^{-P_{\text{tot}}C_{\text{r}} L_{\text{eff}}\nu} d\nu}e^{j\phi\left(f_1,f_2,f,\zeta\right)}\right|^2
\\
&\cdot \left[R^2\left(f-f_1\right) + R\left( f-f_1\right)R\left(  f-f_2\right)\right],
\label{eq:GNmodel_single_pol}
\end{split}
\end{equation}
with 
\begin{equation}
\begin{split}
&R(f) = \frac{1}{\sqrt{2}}\Re \left\{H(f)\right\},
\label{eq:Rf_1pol}
\end{split}
\end{equation}
and where $\frac{8}{9}\to 1$ in the nonlinear transfer function \eqref{eq:response_freq}. Eq. \eqref{eq:GNmodel_single_pol} is an extension of the ISRS GN model to account for the real Raman spectrum for single polarization. The scaling factor for the SPM/XPM contribution in the case of single polarization is 
\begin{equation}
\begin{split}
&R_\text{XPM}\left(\Delta f\right)=\frac{R^2\left(\Delta f\right) + 2R\left( \Delta f\right)R\left(  0\right)+R^2\left( 0\right)}{2}.
\label{eq:XPM_impact_1pol}
\end{split}
\end{equation}
The single polarization case \eqref{eq:GNmodel_single_pol}\eqref{eq:Rf_1pol}\eqref{eq:XPM_impact_1pol} differs from the dual polarization case \eqref{eq:ISRSGNmodel_rho}\eqref{eq:Rf}\eqref{eq:XPM_impact} by more than the usual pre-factor of $2\to \frac{16}{27}$. This is due to the averaging in the random variables, representing the modulation symbols (see Appendix \ref{sec:apx:derivation}b)). In the case of dual polarization, an additional cross-polarization component contributes to the term involving $R\left(f\right)$, resulting in the different functional form between single and dual polarization. In the remainder of this paper, however, only the case of dual polarization is considered. 
\section{Numerical validation}
\label{sec:validation}
In the following, the ISRS GN model \eqref{eq:ISRSGNmodel_rho} is validated by numerical simulations. The numerical simulator aims to solve the generalized Manakov equation \eqref{eq:GME_time} using the split-step Fourier method (SSFM) with parameters listed in Table \ref{tab:parameters}. 
\par 
\ 
To validate the proposed theory, a circular, complex Gaussian constellation is chosen as modulation format to resembles the initial condition in the derivation of the ISRS GN model. The impact of the real Raman spectrum, and the validation of Eq. \eqref{eq:ISRSGNmodel_rho}, is carried out over the C-band (5~THz optical bandwidth) in Section \ref{sec:results_C-band} and over the C+L-band (10~THz optical bandwidth) in Sections \ref{sec:results_CL-band1} and \ref{sec:results_CL-band2}. For each transmission scenario, two different symbol rates were considered; a) 5~GBd with a launch power of -8~dBm per channel and b) 40~GBd with a launch power of 1 dBm per channel. The launch powers were chosen such that both cases exhibit the same power spectral density within a single WDM channel.  
\par
\ 
The SSFM implements the GME as in \eqref{eq:GME_time}. In contrast to previous works \cite{Semrau_2018_tgn, Semrau_2019_aca, Semrau_2019_amf}, ISRS is implemented through the imaginary part of the Raman spectrum and \textit{not} via a signal power profile (i.e. $g\left(z,f\right)=-\alpha$ in \eqref{eq:GME_freq}). The nonlinear response was implemented in the nonlinear step by filtering the optical power over both polarizations with the nonlinear transfer function $H(f)$, as suggested in \cite{Blow_1989_tdo} for solving the GNLSE. The filtering process (convolution operation) in the time domain requires an additional fast Fourier transform (FFT) and inverse fast Fourier transform (iFFT) pair for every nonlinear step, increasing the computational complexity of the SSFM algorithm by around 50\%. The step size was logarithmically distributed with a total of 250000 steps per fiber span. 
\par 
\ 
For validation purposes, only single span transmission was considered, assuming ideal amplification and gain equalization. To ease the comparison of the NLI power between model and simulation, no amplified spontaneous emission (ASE) noise was injected at the amplification stage. At the receiver, digital dispersion compensation, ideal root-raised-cosine (RCC) matched filtering and constellation rotation was carried out. The SNR was then estimated as the ratio between the variance of the transmitted symbols $E[|X|^2]$ and the variance of the noise $\sigma^2$, where $\sigma^2=E[|X-Y|^2]$ and $Y$ represents the received symbols after digital signal processing. In order to improve the simulation accuracy, four different data realizations were simulated and averaged for each transmission.  
\par 
\ 
For the numerical validations, we define the impact of the real Raman spectrum as
\begin{equation}
\begin{split}
&\Delta \eta_1 = \frac{\gamma^2}{\gamma^2_\text{eff}}\frac{\eta_{1}\left(f_i\right)\Big|_{f_r \neq 0}}{\eta_{1}\left(f_i\right)\Big|_{f_r = 0}} =\frac{\gamma^2}{\gamma^2_\text{eff}} \frac{\text{SNR}^{-1}_{\text{NLI,}i}\Big|_{f_r \neq 0}}{\text{SNR}^{-1}_{\text{NLI,}i}\Big|_{f_r = 0}},
\label{eq:fom}
\end{split}
\end{equation}
where $\gamma_\text{eff}=\frac{8+f_r}{8}$ is the effective nonlinearity coefficient (see Sec. \ref{sec:delayed_response}), responsible for the $0.2456$-dB shift in Fig. \ref{fig:XPM_impact}. It should be emphasized that the imaginary part (ISRS) were considered in both cases, regardless of $f_r$. Eq. \eqref{eq:fom} was defined such that it removes the zero'th order impact with respect to dual polarization or the GME (i.e. $\frac{8+f_r}{9} \to \frac{8}{9}\gamma_\text{eff}$ ). This was done for two reasons: 1) In practice, $\gamma_\text{eff}$ might be obtained by fitting experimental measurements using dual polarized signals. As a consequence, the impact of $\gamma_\text{eff}$, as opposed to $\gamma$, might be accounted for by 'default'. 2) The zero'th order impact of the real Raman spectrum on dual polarized signals was discovered in \cite{Antonelli_2016_mon}. To highlight the contribution of this work, we chose to remove the zero'th order impact in this section. In summary, Eq. \eqref{eq:fom} measures the impact of the frequency dependence of the real Raman spectrum, that cannot be modelled and offset by using an effective nonlinearity coefficient.
\par 
\ 
In the next sections, Eq. \eqref{eq:fom} will be evaluated using numerical simulations and the ISRS GN model \eqref{eq:ISRSGNmodel_rho} to validate the theory proposed in this work.
\begin{table}
\renewcommand{\arraystretch}{1.2}
\centering
\caption{System Parameters}
\label{tab:parameters}
  \begin{tabular}{ l | c c }
   \textbf{Parameters} & \textbf{a)} & \textbf{b)} \\  \hline
       Symbol rate [GBd]&  5 & 40\\ \hline
    Channel spacing [GHz]& 5.005 & 40.005 \\ \hline
    Channel Launch Power ($P_{i}$) [dBm]&   -8 & 1\\ \hline
      Reference Wavelength [nm]& \multicolumn{2}{c}{1550} \\ \hline
          Roll-off factor [\%]&  \multicolumn{2}{c}{0.01}  \\ \hline
    Loss ($\alpha$) [dB/km]& \multicolumn{2}{c}{0.16}\\ \hline
    Dispersion ($D$) [ps/nm/km]& \multicolumn{2}{c}{16.4}\\ \hline
     Dispersion slope ($S$) [ps/$\text{nm}^2$/km]& \multicolumn{2}{c}{0.067}\\ \hline
    NL coefficient ($\gamma$) [1/W/km]& \multicolumn{2}{c}{0.104}\\ \hline
   Effective core area [$\mu\text{m}^2$]& \multicolumn{2}{c}{81.8}\\ \hline
    Raman gain slope ($C_{\text{r}}$) [1/W/km/THz]&  \multicolumn{2}{c}{0.0236} \\ \hline
    Raman gain ($C_{\text{r}}\cdot 14$ THz) [1/W/km]& \multicolumn{2}{c}{0.33}\\ \hline
   Number of symbols [$2^{x}$]& 14 & 17  \\ \hline
    Simulation steps per span [$10^6$]& \multicolumn{2}{c}{0.25}  \\ \hline
  \end{tabular}
\end{table}
\subsection{C-band transmission results}
\label{sec:results_C-band}
\begin{figure*}[h!]
   \centering
    \includegraphics[]{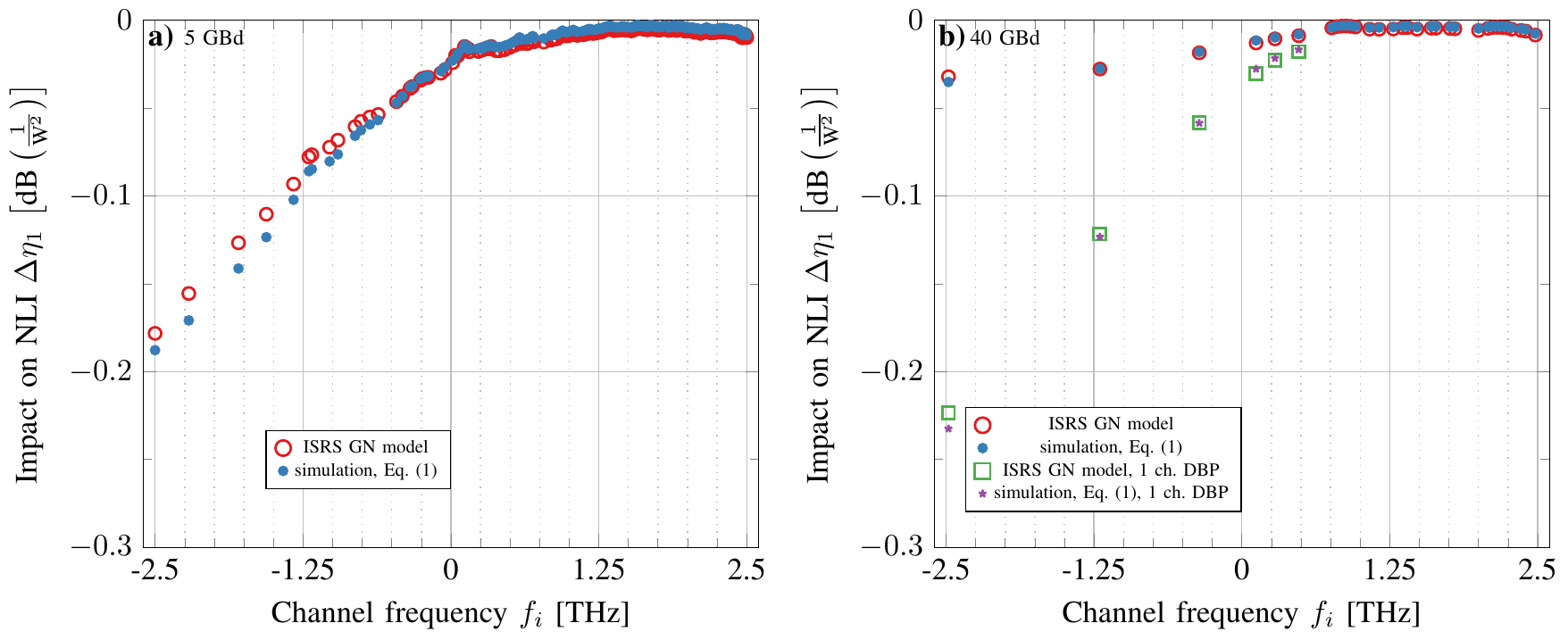}
\caption{The impact of the real Raman spectrum on the nonlinear interference \eqref{eq:fom} as a function of channel frequency over the entire C-band (5 THz) for a) 5~GBd and b) 40~GBd modulated channels. The SPM contribution of the total NLI was removed via single channel DBP in the case of b) 40~GBd modulation. The results were obtained by split-step simulations of Eq. \eqref{eq:GME_time} and numerically integrating the proposed model \eqref{eq:ISRSGNmodel_rho}.}
\label{fig:c_band}
\end{figure*}
In this section, transmission over the entire C-band (5~THz optical bandwidth) is considered. The transmitted channel configurations are spectra that occur in mesh optical network transmission. 501 and 64 channels slots were available for the a) 5GBd and b) 40 GBd case, respectively. The spectral occupancy of the transmitted spectra was around 25\%. 
\par 
\ 
In the following, the chosen transmission spectra are explained in more detail. A prevailing routing and wavelength assignment (RWA) algorithm is $k$-shortest path - first fit (kSP-FF). An incoming demand request for a given node pair, is assigned to the first non-blocking wavelength (i.e. last frequency) slot for the light path setup. If a non-blocking wavelength slot is available on the current path, the light path is established. If no non-blocking wavelength slot is available, the second shortest path between the requested node pair is scanned. If no non-blocking wavelength slot is available for the second shortest path, the third shortest path is scanned which is repeated until the $k$-shortest paths were scanned. kSP-FF results in transmission spectra where high frequency channels have a significantly higher spectral occupation than lower frequency channels (cf. \cite[Fig. 9]{Vincent_2019_sce} ). The reason for this behavior is that the network blocking probability is low at the beginning of the network operation leading to more channels being allocated at higher frequencies. Over time, the blocking probability of a given node pair increases and larger wavelengths need to be scanned to find a non-blocking channel slot. In this work, the used transmission spectra are sampled from an exponential probability distribution to represent signal spectra present in mesh optical networks. In the following, we show that this is one scenario in which the real part of the ISRS has an impact on system performance.
\par 
\ 
The impact of the real part, as in Eq. \eqref{eq:fom}, as a function of channel frequency is shown in Fig. \ref{fig:c_band} for channels modulated at a) 5~GBd and b) 40~GBd. In the case of 40~GBd, the SPM contribution of a few channels was removed via single channel digital back-propagation (DBP). The simulation results represents the results from numerically solving \eqref{eq:GME_time}. The modelling results were obtained by numerically integrating Eq. \eqref{eq:ISRSGNmodel_rho}. The maximum deviation between model and simulation is 0.01~dB and 0.009~dB for 5~GBd and 40~GBd modulation, respectively. The sampling requirements in the numerical integration for 5~GBd modulation is higher with respect to their 40~GBd counterpart. This results in slightly different accuracies between the two symbol rates. 
\par 
\ 
The impact of the real part is always negative, which means that the real Raman spectrum \textit{reduces} the NLI (apart from the $\frac{\gamma^2}{\gamma_\text{eff}^2}$ scaling as shown in Fig. \ref{fig:XPM_impact}). The impact is around 0.006~dB for higher channel frequencies where channels are closely spaced. The impact increases for lower frequency channels were the occupation is more sparse. The impact of the real part is less for 40~GBd channels compared to 5~GBd channels. The reason is that the real part scales the XPM (and FWM) contributions (cf. Fig. \ref{fig:XPM_impact}). 40~GBd channels exhibit a larger SPM contribution which is unchanged by the real Raman spectrum. In the case of 5~GBd channels, the XPM fraction of the total NLI is higher yielding a larger impact of the real Raman spectrum. This effect becomes particularly evident when single channel DBP is applied to 40~GBd channels, completely removing the SPM contribution. For the channel located at $f_i=2.5$~THz, the real Raman spectrum decreases the NLI by 0.224~dB.   
\par 
\ 
The analysis shows that the impact of the real Raman spectrum is strongly dependent on the channel occupancy across the optical spectrum. The impact is shown to be larger for channels that exhibit a low relative SPM contribution. Additionally, the analysis shows that the impact of the real Raman spectrum is not only a property of ultra-wideband transmission and is also significant for C-band transmission systems. 

\subsection{C+L-band transmission results I}
\label{sec:results_CL-band1}
\begin{figure*}[h!]
   \centering
    \includegraphics[]{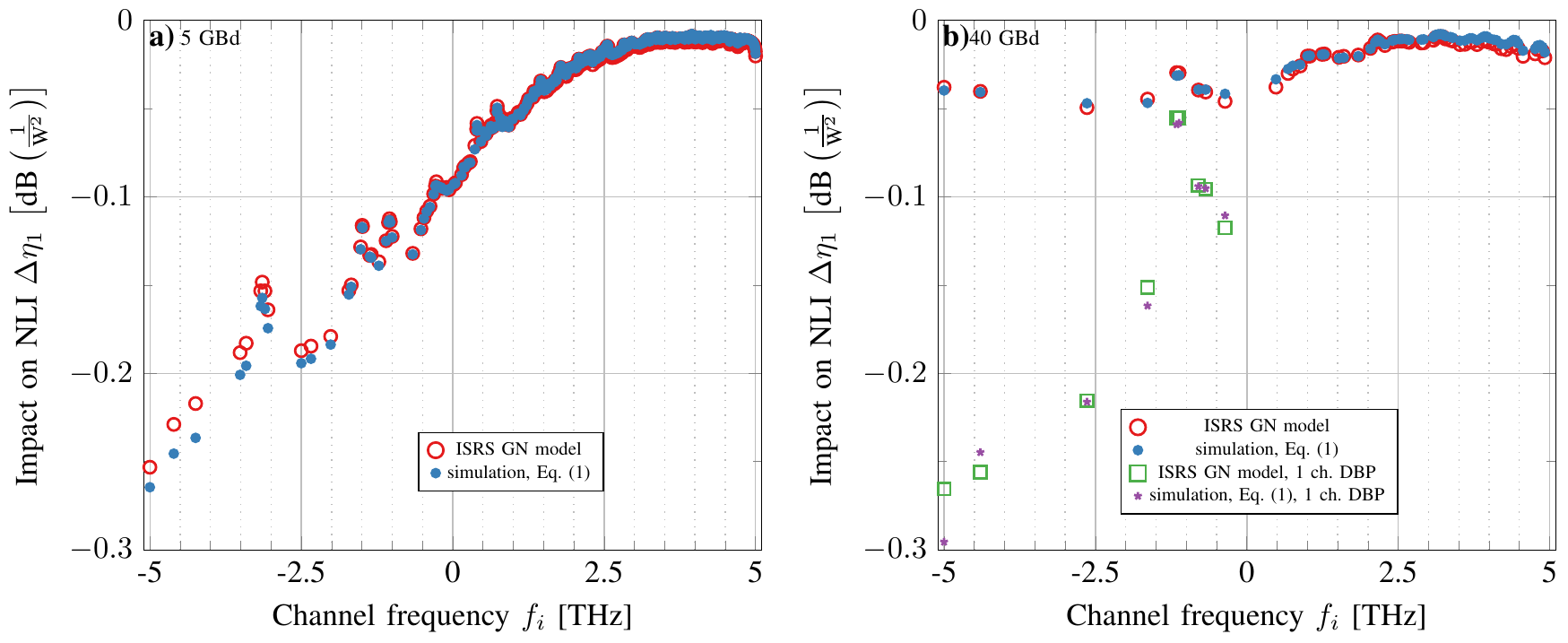}
\caption{The impact of the real Raman spectrum on the nonlinear interference \eqref{eq:fom} as a function of channel frequency over the entire CL-band (10 THz) for a) 5~GBd and b) 40~GBd modulated channels. The SPM contribution of the total NLI was removed via single channel DBP in the case of b) 40~GBd modulation. The results were obtained by split-step simulations of Eq. \eqref{eq:GME_time} and numerically integrating the proposed model \eqref{eq:ISRSGNmodel_rho}.}
\label{fig:cl_band}
\end{figure*}
In this section, the numerical validation is extended to a C+L band transmission system with the same transmission spectra characteristics as in Sec. \ref{sec:results_C-band}. 
\par 
\ 
The impact of the real part, as in Eq. \eqref{eq:fom}, as a function of channel frequency is shown in Fig. \ref{fig:cl_band} for channels modulated at a) 5~GBd and b) 40~GBd. Again, the SPM contribution of a few 40~GBd channels was compensated through single channel DBP. The maximum mismatch between simulation and model is 0.01~dB and 0.03~dB for a) 5~GBd and b) 40~GBd, respectively. 
\par 
\ 
The maximum impact of the real Raman spectrum is around 0.267 and 0.296~dB for the lowest frequency channel located at -5~THz. Comparing the C+L-band transmission case with the C-band case, there are larger XPM (and FWM) contributions that exhibit larger frequency spacings which are scaled stronger by the real Raman spectrum (cf. Fig. \ref{fig:XPM_impact}).
\par 
\
The C and C+L-band transmission results shown in Fig. \ref{fig:c_band} and \ref{fig:cl_band} indicate the real Raman spectrum is relevant in network transmission scenarios for low symbol rate channels and high symbol rate channels that employ some form of narrow-band nonlinearity compensation. 
\subsection{C+L-band transmission results II}
\label{sec:results_CL-band2}
\begin{figure*}[h!]
   \centering
    \includegraphics[]{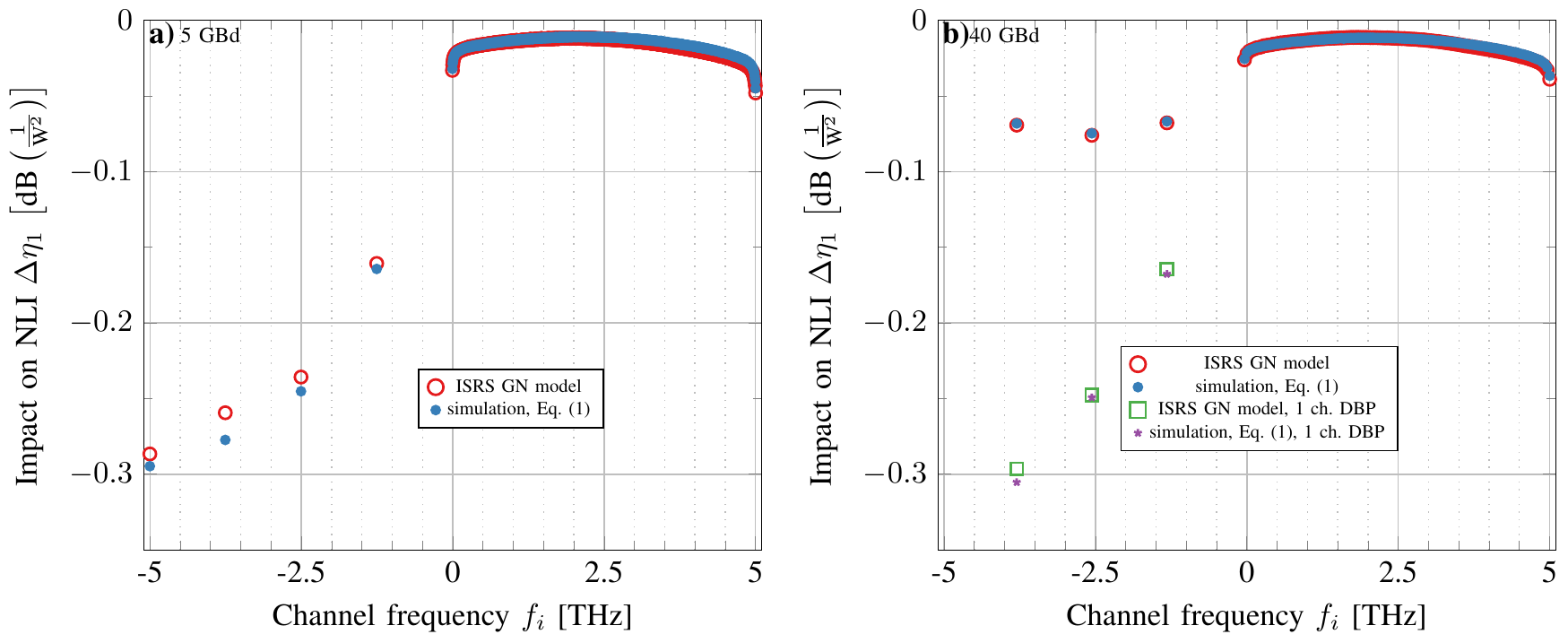}
\caption{The impact of the real Raman spectrum on the nonlinear interference \eqref{eq:fom} as a function of channel frequency over the entire CL-band (10 THz) for a) 5~GBd and b) 40~GBd modulated channels. The transmission specta represent a fully occupied C-band and a few channels lit up in the L-band. The SPM contribution of the total NLI was removed via single channel DBP in the case of b) 40~GBd modulation. The results were obtained by split-step simulations of Eq. \eqref{eq:GME_time} and numerically integrating the proposed model \eqref{eq:ISRSGNmodel_rho}.}
\label{fig:cl_band_full_c}
\end{figure*}
In this section, the numerical validation is carried out for a C+L-band transmission, similar to Sec. \ref{sec:results_CL-band1}, but with a different transmission spectrum. It is assumed that the entire C-band is occupied and only a few channels are established in the L-band. This scenario may occur when the C-band is exhausted and operators decide to extend the transmission window to the L-band. New demands are allocated in the L-band, while the C-band exhibits a high spectral occupancy. In this example, the C-band is fully occupied, whereas 4 and 3 channels are present in the L-band for a) 5 GBd and b) 40 GBd, respectively.
\par 
\ 
The impact of the real part as a function of channel frequency is shown in Fig. \ref{fig:cl_band_full_c} for channels modulated at a) 5~GBd and b) 40~GBd. The SPM contribution of a few 40~GBd channels was removed through single channel DBP. The maximum mismatch between simulation and model is 0.008~dB and 0.003~dB for a) 5~GBd and b) 40~GBd, respectively.  
\par 
\ 
The maximum impact of the real Raman spectrum within the C-band is 0.05 dB, which is because most XPM (and FMW) contributions for those channels originate from closely spaced channels, that are not strongly affected by the real Raman spectrum (cf. Fig \ref{fig:XPM_impact}). On the other hand, the maximum impact of the real Raman spectrum on the L-band channels is 0.295 and 0.3~dB for a) 5~GBd and 40~GBd, respectively. This is because most of the NLI in the L-band originates from the C-band that has a large frequency separation and is, therefore, scaled by the real Raman spectrum. 
\par 
\ 
The study in this section indicates that the real Raman spectrum is negligible for fully occupied transmission scenarios. However, for scenarios where some bands exhibit higher spectral occupancies than other transmission bands, the real Raman spectrum becomes relevant. Similar to Sec. \ref{sec:results_CL-band2}, the real Raman spectrum has a larger impact for low symbol rate channels or high symbol rate channels that employ some form of narrow-band nonlinearity compensation. 
\par 
\ 
In conclusion, the extension of ISRS GN model that includes the real Raman spectrum \eqref{eq:ISRSGNmodel_rho}\eqref{eq:ISRSGNmodel_Cr} exhibits excellent accuracy compared to numerical simulations, validating the theory proposed in this work. The derived model can be used in the design and operation of optical communication links and networks. 
\section{Conclusion}
For the first time, the impact of the real Raman spectrum on the nonlinear interference of coherent optical transmission systems was numerically and analytically investigated. Starting from a generalized Manakov equation, an extension of the ISRS GN model was derived that accounts for the real part of the complex valued Raman spectrum. Enabled by a closed-form approximation of the extended ISRS GN model, it is analytically shown that the real Raman spectrum scales the NLI, in particular the XPM terms, dependent on the frequency separation of the interacting frequencies. A novel analytical approximation of the real Raman spectrum is proposed enabling fully analytical evaluations of the real Raman spectrum. The newly derived model was numerically validated in C- and C+L-band transmission systems. 
\par 
\ 
Additionally, it is shown that the real Raman spectrum is relevant for transmission spectra that have a varying degree in spectral occupancy across the optical spectrum, such as in mesh optical network transmission. This is particularly true for channels that exhibit low SPM contributions, such as low symbol rate channels or systems that deploy narrow-band nonlinearity compensation (e.g. single channel DBP). It was demonstrated that the real Raman spectrum can be neglected in fully populated transmission systems, where the NLI is dominated from closely spaced channels.

\appendices

\section{Relation of the GME \eqref{eq:GME_time} to \cite[Eq. (103)]{Antonelli_2016_mon}.}
\label{apx:sec:GME_and_antonelli}
In this section, it is shown that the GME \eqref{eq:GME_time} is similar to the result in \cite[Eq. (103)]{Antonelli_2016_mon} for dual polarisation and \cite[Eq. (12)]{Blow_1989_tdo} for single polarisation. For this purpose, the delayed electrical field is written as an infinite Taylor series for both polarizations as
\begin{equation}
\begin{split}
&\left|E_\text{x}\left(t-\tau\right)\right|^2=\sum_0^{\infty} \frac{\left(-\tau\right)^n}{n!}\frac{\partial^n}{\partial t^n}\left|E_\text{x}\left(t\right)\right|^2, \\ 
&\left|E_\text{y}\left(t-\tau\right)\right|^2=\sum_0^{\infty} \frac{\left(-\tau\right)^n}{n!}\frac{\partial^n}{\partial t^n}\left|E_\text{y}\left(t\right)\right|^2.
\label{eq:field_taylor}
\end{split}
\end{equation}
Inserting the Taylor series description of the delayed field \eqref{eq:field_taylor} into the GME \eqref{eq:GME_time} yields 
\begin{equation}
\begin{split}
&\frac{\partial}{\partial z}E_\text{x}\left(z,t\right)=\left(-\frac{\alpha}{2}-j\frac{\beta_2}{2}\frac{\partial^2}{\partial t^2}+\frac{\beta_3}{6}\frac{\partial^3}{\partial t^3}\right)E_\text{x}\left(z,t\right) \\
&+j\gamma E_\text{x}\left(z,t\right)\sum_0^{\infty} \frac{1}{n!}\int h\left(\tau\right)\left(-\tau\right)^nd\tau\\
&\cdot \frac{\partial^n}{\partial t^n}\left[\left|E_\text{x}\left(z,t\right)\right|^2+\left|E_\text{y}\left(z,t\right)\right|^2\right]
\label{appendixA:eq:GME1}
\end{split}
\end{equation}
To relate Eq. \eqref{appendixA:eq:GME1} to the nonlinear transfer function $H\left(f\right)$,
the following identity for the $n$'th moment of a function is used 
\begin{equation}
\begin{split}
&\int \left(-\tau\right)^n h\left(\tau\right) d\tau = \left(\frac{j}{2\pi}\right)^n\frac{\partial^n}{\partial f^n}H\left(f\right)\bigg|_{f=0},
\end{split}
\end{equation}
to obtain
\begin{equation}
\begin{split}
&\frac{\partial}{\partial z}E_\text{x}\left(z,t\right)=\left(-\frac{\alpha}{2}-j\frac{\beta_2}{2}\frac{\partial^2}{\partial t^2}+\frac{\beta_3}{6}\frac{\partial^3}{\partial t^3}\right)E_\text{x}\left(z,t\right) \\
&+j\gamma E_\text{x}\left(z,t\right)\sum_0^{\infty} \frac{j^n}{\left(2\pi\right)^nn!}\frac{\partial^n}{\partial f^n}H\left(f\right)\bigg|_{f=0}\\
&\cdot \frac{\partial^n}{\partial t^n}\left[\left|E_\text{x}\left(z,t\right)\right|^2+[\left|E_\text{y}\left(z,t\right)\right|^2\right].
\label{appendixA:eq:GME_taylor}
\end{split}
\end{equation}
The real Raman spectrum $n_r\left(f\right)$ only consists of even orders, while the imaginary part $g_r\left(f\right)$ only consists of odd orders. Truncating \eqref{appendixA:eq:GME_taylor} to first order and using the normalization as in \eqref{eq:fr} (i.e. $n_r\left(0\right)=1$) yields
\begin{equation}
\begin{split}
&\frac{\partial}{\partial z}E_\text{x}\left(z,t\right)= \left(-\frac{\alpha}{2}-j\frac{\beta_2}{2}\frac{\partial^2}{\partial t^2}+\frac{\beta_3}{6}\frac{\partial^3}{\partial t^3}\right)E_\text{x}\left(z,t\right) \\
&+j\gamma E_\text{x}\left(z,t\right) \left[\left|E_\text{x}\left(z,t\right)\right|^2+\left|E_\text{y}\left(z,t\right)\right|^2\right]\\ &\cdot\left[\frac{8}{9}\left(1-f_r\right)+f_r-T_r\frac{\partial}{\partial t}\right]
\label{appendixA:eq:GME_2ndorder}
\end{split}
\end{equation}
with the Raman time constant
\begin{equation}
\begin{split}
T_r = \frac{1}{2\pi}\frac{\partial}{\partial f}g_r\left(f\right)\bigg|_{f=0}= \frac{\lambda_0\tilde{C}_r}{8\pi^2 n_2}.
\end{split}
\end{equation}
Eq. \eqref{appendixA:eq:GME_2ndorder} is identical to the result in \cite[Eq. (103)]{Antonelli_2016_mon} (in Dirac notation) which is the first-order approximation of \eqref{eq:GME_time} with respect to the delayed electrical field. The Raman time constant is estimated to $T_r=3.6$~fs, using $n_2=2.1\cdot 10^{-20}$~$\frac{\text{m}^2}{\text{W}}$ and the Raman gain spectrum as in Fig. \ref{fig:raman_spectrum}. A value of $n_2=2.6\cdot 10^{-20}$~$\frac{\text{m}^2}{\text{W}}$ yields a time constant as $T_r=2.94$~fs, consistent with \cite{Agrawal_2012_nfo}.
\section{Deriving the field description of the Raman gain equations from the GME \eqref{eq:GME_time}}
\label{sec:apx:GMEtoODE}
In this section, it is shown that the ISRS power transfer resulting from the imaginary Raman spectrum in the GME \eqref{eq:GME_time} resembles the one obtained from the Raman gain equations. The Raman gain equations is a system of coupled differential equations that describe the effect of ISRS in terms of optical power \cite[Eq.~(3)]{Tariq_1993_acm}. The GME including the imaginary Raman spectrum, on the other hand, describes ISRS in terms of the electrical field, including temporal gain dynamics and interactions between dispersion and Kerr nonlinearity. In the following, the GME and its normalization \eqref{eq:spec_prefactors} are confirmed by showing that the GME reduces to the Raman gain equations in the absence of dispersion and instantaneous Kerr nonlinearity. Additionally, a polarization sensitive field description of the Raman gain equations is derived that might be useful in developing techniques to mitigate ISRS. 
\par 
\
For simplicity, we consider a transmitted signal consisting of only two continuous waves with powers $P_{\text{x,}0}=\left|E_{\text{x,}0}\right|^2$ and $P_{\text{x,}k}=\left|E_{\text{x,}k}\right|^2$ for both polarizations at (relative) frequencies $0$ and $f_k$, respectively. The waveform at the transmitter is then given by 
\begin{align}
\begin{split}
E_\text{x}\left(f\right)= \left[E_{\text{x,}0} \delta\left(f\right) + E_{\text{x,}k} \delta\left(f-f_k\right)\right],\\
E_\text{y}\left(f\right)= \left[E_{\text{y,}0} \delta\left(f\right) + E_{\text{y,}k} \delta\left(f-f_k\right)\right],
\label{appendixB:eq:tx_signal}
\end{split}
\end{align}
where $E_{\text{x,}0}\left(z\right)$, $E_{\text{y,}0}\left(z\right)$, $E_{\text{x,}k}\left(z\right)$ and $E_{\text{y,}k}\left(z\right)$ are complex valued and distance-dependent, which is suppressed for notational convenience. Substituting \eqref{appendixB:eq:tx_signal} in \eqref{eq:GME_time} and neglecting dispersion, the instantaneous part of the nonlinear response and the real part of the Raman response yields
\begin{equation}
\begin{split}
&\frac{\partial}{\partial z}\left[E_{\text{x,}0} \delta\left(f\right) + E_{\text{x,}k} \delta\left(f-f_k\right)\right]=\\
&-\frac{\alpha}{2}\left[E_{\text{x,}0} \delta\left(f\right) + E_{\text{x,}k} \delta\left(f-f_k\right)\right]\\
&-\int df_1\int df_2\left[E_{\text{x,}0} \delta\left(f-f_2\right) + E_{\text{x,}k} \delta\left(f-f_2-f_k\right)\right]\\
&f_r\gamma g_r(f_2)\left[E_{\text{x,}0} \delta\left(f_1\right) + E_{\text{x,}k} \delta\left(f_1-f_k\right)\right]\\
&\left[E^{*}_{\text{x,}0} \delta\left(f_1-f_2\right) + E^{*}_{\text{x,}k} \delta\left(f_1-f_2-f_k\right)\right]\\
&- \int df_1\int df_2\left[E_{\text{x,}0} \delta\left(f-f_2\right) + E_{\text{x,}k} \delta\left(f-f_2-f_k\right)\right]\\
&f_r\gamma g_r(f_2)\left[E_{\text{y,}0} \delta\left(f_1\right) + E_{\text{y,}k} \delta\left(f_1-f_k\right)\right]\\
&\left[E^{*}_{\text{y,}0} \delta\left(f_1-f_2\right) + E^{*}_{\text{y,}k} \delta\left(f_1-f_2-f_k\right)\right].
\label{appendixb:eq:basf}
\end{split}
\end{equation}
The convolution integrals are only non-zero for two frequency combinations that are $\left\{f_1=0 \ \text{and} \ f_2=-f_k\right\}$ and $\left\{f_1=f_k \ \text{and} \ f_2=f_k\right\}$. Exploiting the fact that $g_r(-f_k)=-g_r(f_k)$, Eq. \eqref{appendixb:eq:basf} reduces to
\begin{equation}
\begin{split}
&\frac{\partial}{\partial z}\left[E_{\text{x,}0} \delta\left(f\right) + E_{\text{x,}k} \delta\left(f-f_k\right)\right]=\\
&-\frac{\alpha}{2}\left[E_{\text{x,}0} \delta\left(f\right) + E_{\text{x,}k} \delta\left(f-f_k\right)\right]\\
&+f_r\gamma g_r(f_k)\left[E_{\text{x,}0}E^{*}_{\text{x,}k}E_{\text{x,}k} \delta\left(f\right)- E_{\text{x,}0}E_{\text{x,}k}E^{*}_{\text{x,}0} \delta\left(f-f_k\right) \right]\\
&+f_r\gamma g_r(f_k)\left[E_{\text{y,}0}E^{*}_{\text{y,}k}E_{\text{x,}k} \delta\left(f\right)- E_{\text{x,}0}E_{\text{y,}k}E^{*}_{\text{y,}0} \delta\left(f-f_k\right) \right]. 
\label{appendixB:eq:idnasiddada}
\end{split}
\end{equation}
As the Dirac delta functions in \eqref{appendixB:eq:idnasiddada} are orthogonal, they can be written into two separate, coupled equations
\begin{equation}
\begin{split}
&\frac{\partial}{\partial z}E_{\text{x,}0} =-\frac{\alpha}{2}E_{\text{x,}0} +f_r\gamma g_r(f_k)E_{\text{x,}0}P_{\text{x,}k} \\
&+f_r\gamma g_r(f_k)E_{\text{y,}0}E^{*}_{\text{y,}k}E_{\text{x,}k} ,\\
&\frac{\partial}{\partial z}E_{\text{x,}k}=-\frac{\alpha}{2}E_{\text{x,}k}-f_r\gamma g_r(f_k) E_{\text{x,}k}P_{\text{x,}0}\\
&-f_r\gamma g_r(f_k) E_{\text{x,}0}E_{\text{y,}k}E^{*}_{\text{y,}0}.
\label{appendixB:eq:idnasid}
\end{split}
\end{equation}
Eq. \eqref{appendixB:eq:idnasid} is then transformed to the optical power using $\frac{\partial P}{\partial z} = \frac{\partial E^{*}}{\partial z}E+\frac{\partial E}{\partial z}E^{*}$ resulting in
\begin{equation}
\begin{split}
&\frac{\partial}{\partial z}P_{\text{x,}0} =-\alpha P_{\text{x,}0} +\frac{\tilde{g}_r(f_k)}{A_\text{eff}}P_{\text{x,}0}P_{\text{x,}k} \\
&+\frac{\tilde{g}_r(f_k)}{2A_\text{eff}}\left[E^{*}_{\text{y,}0}E_{\text{x,}0}E_{\text{y,}k}E^{*}_{\text{x,}k}+E_{\text{y,}0}E^{*}_{\text{x,}0}E^{*}_{\text{y,}k}E_{\text{x,}k} \right] 
\\
&\frac{\partial}{\partial z}P_{\text{x,}k} =-\alpha P_{\text{x,}k}-\frac{\tilde{g}_r(f_k)}{A_\text{eff}} P_{\text{x,}k}P_{\text{x,}0}\\
&-\frac{\tilde{g}_r(f_k)}{2A_\text{eff}}\left[E^{*}_{\text{x,}0}E_{\text{x,}k}E^{*}_{\text{y,}k}E_{\text{y,}0}+E^{*}_{\text{x,}k}E_{\text{x,}0}E_{\text{y,}k}E^{*}_{\text{y,}0}\right].
\label{apx:eq:adsadsadasdada}
\end{split}
\end{equation}
Assuming that the signal in the respective polarization states are different, the cross-polarization terms involving the complex field envelope in \eqref{apx:eq:adsadsadasdada} are approximately zero. Eq. \eqref{apx:eq:adsadsadasdada} furthermore indiciates that the ISRS power transfer is \textit{doubled (in log-scale)}, when x and y polarization contain the same signal. On the contrary, the ISRS power transfer is \textit{removed} when $E^{*}_{\text{y,}0}E_{\text{x,}0}E_{\text{y,}k}E^{*}_{\text{x,}k}+E_{\text{y,}0}E^{*}_{\text{x,}0}E^{*}_{\text{y,}k}E_{\text{x,}k} =-2P_{\text{x,}0}P_{\text{x,}k}$ (e.g. $E_{\text{y,}0}=E_{\text{x,}0}$ and $E_{\text{y,}k}=-E_{\text{x,}k}$). The field description of the Raman gain equations in \eqref{apx:eq:adsadsadasdada} can be potentially used to derive ISRS mitigation techniques.
\par 
\ 
Finally, the optical power across both polarizations is obtained by $\frac{\partial}{\partial z}P_0=\frac{\partial}{\partial z}P_{\text{x,}0}+\frac{\partial}{\partial z}P_{\text{y,}0}$ yielding
\begin{equation}
\begin{split}
&\frac{\partial}{\partial z}P_0 =-\alpha P_0 +\frac{1}{A_\text{eff}}\tilde{g}_r(f_k)\left(P_{\text{x,}0}P_{\text{x,}k}+P_{\text{y,}0}P_{\text{y,}k}\right) \\
&\frac{\partial}{\partial z}P_k =-\alpha P_k -\frac{1}{A_\text{eff}} \tilde{g}_r(f_k) \left(P_{\text{x,}0}P_{\text{x,}k}+P_{\text{y,}0}P_{\text{y,}k}\right).
\end{split}
\end{equation}
For totally scrambled polarization states, we have $P_{\text{x,}0}\approx \frac{1}{2}P_0$ and $P_{\text{x,}k}\approx \frac{1}{2}P_k$ and obtain  
\begin{equation}
\begin{split}
&\frac{\partial}{\partial z}P_0 =-\alpha P_0 +\frac{1}{2A_\text{eff}}\tilde{g}_r(f_k)P_0P_k \\
&\frac{\partial}{\partial z}P_k =-\alpha P_k -\frac{1}{2A_\text{eff}} \tilde{g}_r(f_k) P_0P_k.
\label{appendixb:eq:diadia}
\end{split}
\end{equation}
Eq. \eqref{appendixb:eq:diadia} is identical to the Raman gain equations \cite[Eq.~(3)]{Tariq_1993_acm} for two co-propagating continuous waves (cw case). The factor $\frac{1}{2}$ is typically referred to as the polarization averaging factor. Note that \eqref{appendixb:eq:diadia} does not include the ratio of the photon energies which can, however, be neglected in most cases as $\frac{c+f_i\lambda_0}{c+f_k\lambda_0}\approx 1$, with $c$ being the speed of light in vacuum. The reader is reminded that $f_i$ and $f_k$ are relative frequencies centered at $\frac{c}{\lambda_0}$.
\section{Analytical description of the real Raman spectrum $\tilde{n}_r\left(f\right)$}
\label{sec:analytical_real_part}
In this section, the real part of the Raman gain spectrum $\tilde{n}_r\left(f\right)$ is derived from its imaginary part $\tilde{g}_r\left(f\right)$. For this purpose, we exploit the linearity of the KK-relations \eqref{eq:kk_relations} and make use of the identity 
\begin{equation}
\begin{split}
\tilde{A}_r\text{cos}\left(\tilde{\omega}_r f\right)=\mathcal{H}\left\{\tilde{A}_r\text{sin}\left(\tilde{\omega}_r f\right)\right\}.
\label{appendixB:kk_sine}
\end{split}
\end{equation}
Using \eqref{eq:kk_relations} and \eqref{appendixB:kk_sine}, the real part Raman spectrum is 
\begin{equation}
\begin{split}
&\tilde{n}_r\left(f\right) = \frac{\tilde{C}_r}{\pi}\text{p.v.}\int_{-\frac{\tilde{B}_r}{2}}^{\frac{\tilde{B}_r}{2}} df'\frac{f'}{f'-f}+\tilde{A}_r\text{cos}\left(\tilde{\omega}_r f\right) \\
&=\frac{\tilde{C}_r}{\pi}\lim_{\epsilon\to 0}\left[\int_{-\frac{\tilde{B}_r}{2}}^{f-\epsilon} df'\frac{f'}{f'-f} +\int_{f+\epsilon}^{\frac{\tilde{B}_r}{2}} df'\frac{f'}{f'-f} \right] \\
&+\tilde{A}_r\text{cos}\left(\tilde{\omega}_r f\right) \\
&=\frac{\tilde{C}_r}{\pi}\lim_{\epsilon\to 0}\left[  f\log\left(\frac{\tilde{B}_r}{2}-f\right)-f\log\left(-\frac{\tilde{B}_r}{2}-f\right)\right.\\
&\left.+\tilde{B}_r-2\epsilon \right]+\tilde{A}_r\text{cos}\left(\tilde{\omega}_r f\right)\\
&=\frac{\tilde{C}_r}{\pi}\left[ f\log\left(\left|\frac{2f-\tilde{B}_r}{2f+\tilde{B}_r}\right|\right)+B_r \right]+\tilde{A}_r\text{cos}\left(\tilde{\omega}_r f\right).
\label{appendixb:eq:result}
\end{split}
\end{equation}
Eq. \eqref{appendixb:eq:result} yields the real part of the complex Raman spectrum as in \eqref{eq:real_part_analytical}.
\section{Extension of the ISRS GN model to include the real Raman spectrum}
\label{sec:apx:derivation}
In this section, the extension of the ISRS GN model is derived, to account for the real part of the complex valued Raman spectrum. First, the case of single polarisation is derived in Sec. \ref{appendix:sec:single_pol}, resulting in Eq. \eqref{eq:GNmodel_single_pol}. In Sec. \ref{appendix:sec:dual_pol}, the case of dual polarisation is addressed, resulting in the extension of the ISRS GN model for dual polarisation \eqref{eq:ISRSGNmodel_rho}\eqref{eq:ISRSGNmodel_Cr}.
\subsection{Single polarized signals}
\label{appendix:sec:single_pol}
In this Section, Eq. \eqref{eq:GNmodel_single_pol} is derived. The impact of ISRS, the imaginary part of $H_r\left(f\right)$, is modelled by a generic frequency and distance dependent gain coefficient $g\left(z,f\right)$, as carried out in \cite[Appendix A]{Semrau_2018_tgn}. Considering the case of single polarisation ($E_\text{y}\left(f\right)=0$) in Eq. \eqref{eq:GME_freq} yields 
\begin{equation}
\begin{split}
\frac{\partial E_\text{x}\left(z,f\right)}{\partial z} &=\widetilde{\Gamma}(z,f)E_\text{x}\left(z,f\right) \\
&+j\gamma E_\text{x}\left(z,f\right)*\left\{H\left(f\right)\left[E_\text{x}(z,f)*E_\text{x}^{*}(z,-f)\right]\right\}.
\label{der:eq:NLSE}
\end{split}
\end{equation}
Eq. \eqref{der:eq:NLSE} is then analytically solved using a first-order regular perturbation  series with respect to the nonlinearity coefficient $\gamma$ as
\begin{equation}
\begin{split}
E_\text{x}(z,f) = E_\text{x}^{\left(0\right)}(z,f)+\gamma E_\text{x}^{\left(1\right)}(z,f).
\label{apx:eq:rp}
\end{split}
\end{equation}
Inserting the RP series \eqref{apx:eq:rp} in \eqref{der:eq:NLSE} yields the zeroth order solution as
\begin{equation}
\begin{split}
E_\text{x}^{\left(0\right)}(z,f) =E_\text{x}(0,f)\cdot e^{\Gamma \left(z,f\right)},
\label{apx:eq:rp1}
\end{split}
\end{equation}
with $\Gamma \left(z,f\right)=\int_0^z\widetilde{\Gamma}\left(\zeta,f\right)d\zeta$ and the first order solution as
\begin{equation}
\begin{split}
E_\text{x}^{(1)}(z,f) =  e^{\Gamma\left(z,f\right)}\int_0^z\frac{Q(\zeta,f)}{e^{\Gamma\left(\zeta,f\right)}}d\zeta,
\label{apx:eq:rp2}
\end{split}
\end{equation}
with 
\begin{equation}
\begin{split}
&Q(z,f)=j \int \int df_1df_2 \\
& E_\text{x}^{(0)}\left(f-f_2\right)H\left(f_2\right)E_\text{x}^{(0)}(f_1)E_\text{x}^{(0)*}\left(f_1-f_2\right),
\label{apx:eq:rp3}
\end{split}
\end{equation}
Eqs. \eqref{apx:eq:rp1}\eqref{apx:eq:rp2}\eqref{apx:eq:rp3} are the generic first-order approximation of the actual solution of the generalized NLSE \eqref{der:eq:NLSE}. However, to calculate the nonlinear perturbation caused by the first-order term, an initial condition $E_\text{x}(0,f)$ is needed. For GN model approaches, the initial condition is an infinitely dense frequency comb, normalized to the PSD of the actual transmitted signal, where every frequency component carries a complex, circular, Gaussian distributed symbol \cite[Eq. (13)]{Carena_2012_mot} as
\begin{equation}
\begin{split}
E_\text{x}^{(0)}(0,f) &= \sqrt[]{f_0G_{\text{Tx}}(f)} \sum_{n=-\infty}^{\infty}\xi_n \delta\left(f-nf_0\right),\\
E_\text{x}^{(1)}(0,f) &= 0,
\label{apx:eq:signal}
\end{split}
\end{equation}
where $G_{\text{Tx}}(f)$ is the power spectral density of the input signal, $\xi_n$ is a complex circular Gaussian distributed random variable and $T_0=f_0^{-1}$ is the period of the signal. For notational brevity, we write $nf_0$ as $f_n$ and $\sum_{n=\infty}^{\infty}$ as $\sum_{\forall n}$ for the remainder of this derivation. Inserting the initial condition \eqref{apx:eq:signal} into Eqs. \eqref{apx:eq:rp1}\eqref{apx:eq:rp2}\eqref{apx:eq:rp3} yields  
\begin{equation}
\begin{split}
&Q(z,f)= jf_0^{\frac{3}{2}}  \sqrt[]{G_{\text{Tx}}(f-f_m + f_l)G_{\text{Tx}}(f_m)G_{\text{Tx}}( f_l)}\\
&\sum_{\forall n}\sum_{\forall m}\sum_{\forall l}\xi_n\xi_m\xi^*_l \delta\left(f- f_m + f_l-f_n\right) \\
& H\left( f_m - f_l\right)e^{\Gamma \left(z,f- f_m +f_l\right)+\Gamma \left(z,f_m\right)+\Gamma^* \left(z,f_l\right)}
\label{apx:eq:QQ}
\end{split}
\end{equation}
As shown in \cite[Ch. \rom{4}.B and \rom{4}.D]{Poggiolini_2012_ada}, only non-degenerate frequency triplets in \eqref{apx:eq:QQ} contribute to the nonlinear interference power to first order. Similar to \cite{Poggiolini_2012_ada, Semrau_2018_tgn}, we define the triplets of non-degenerate frequency components as 
\begin{equation}
\begin{split}
A_i = \left\{(m,n,k):[m-l+n] = i\:\text{and}\:[m\neq l \:\text{or}\: n\neq l]\right\},
\end{split}
\end{equation}
and rewrite \eqref{apx:eq:QQ} as 
\begin{equation}
\begin{split}
&Q(z,f)= \sum_{i=-\infty}^{\infty}\delta\left(f- f_i\right)jf_0^{\frac{3}{2}} \\
& \sum_{\forall (m,n,l) \in A_i}\sqrt[]{G_{\text{Tx}}(f_n)G_{\text{Tx}}(f_m)G_{\text{Tx}}( f_l)}\xi_n\xi_m\xi^*_l H\left( f_m - f_l\right)\\
&e^{\Gamma \left(z,f_n\right)+\Gamma \left(z,f_m\right)+\Gamma^* \left(z,f_l\right)}.
\label{apx:eq:QQQ}
\end{split}
\end{equation}
Inserting \eqref{apx:eq:QQQ} in \eqref{apx:eq:rp2}, yields the first order nonlinear perturbation as
\begin{equation}
\begin{split}
&E_\text{x}^{(1)}(z,f) =  e^{\Gamma\left(z,f\right)} \sum_{\forall i}\delta\left(f- f_i\right)jf_0^{\frac{3}{2}}\sum_{\forall (m,n,l) \in A_i} \\ 
&\cdot\sqrt[]{G_{\text{Tx}}(f_n)G_{\text{Tx}}(f_m)G_{\text{Tx}}( f_l)}\xi_n\xi_m\xi^*_l H\left( f_m - f_l\right)\\
&\mu\left(f_n,f_m,f_l\right),
\end{split}
\end{equation}
with
\begin{equation}
\begin{split}
&\mu\left(f_n,f_m,f_l\right)=\\
&\int_0^z e^{\Gamma \left(z,f_n\right)+\Gamma \left(z,f_m\right)+\Gamma^* \left(z,f_l\right)-\Gamma\left(\zeta,f_m-f_l+f_n\right)}d\zeta.
\end{split}
\end{equation}
The PSD of the nonlinear interference can then be calculated as
\begin{equation}
\begin{split}
&G(f) =\gamma^2\textbf{E}\left[\left|E_\text{x}^{(1)}(z,f)\right|^2\right]\\
&=\gamma^2\sum_{\forall i}\delta\left(f- f_i\right)\sum_{\forall i'}\delta\left(f- f_{i'}\right)\\
& \cdot  e^{2\text{Re}\left[\Gamma\left(z,f\right)\right]} f_0^{3}  \sum_{\forall (m,n,l) \in A_i} \sum_{\forall (m',n',l') \in A_i'}\\
&\cdot \textbf{E}\left[\xi_n\xi_m\xi^*_l \xi^*_{n'}\xi^*_{m'}\xi_{l'}\right]H\left( f_m - f_l\right)H\left( f_{m'} - f_{l'}\right)\\
&\cdot\sqrt[]{G_{\text{Tx}}(f_n)G_{\text{Tx}}(f_m)G_{\text{Tx}}( f_l)G_{\text{Tx}}(f_{n'})G_{\text{Tx}}(f_{m'})G_{\text{Tx}}( f_{l'})}\\
&\cdot\mu\left(f_n,f_m,f_l\right)\mu^*\left(f_{n'},f_{m'},f_{l'}\right),
\label{apx:eq:isahdia}
\end{split}
\end{equation}
where $\textbf{E}\left[x\right]$ denotes the expectation operator. The expectation in \eqref{apx:eq:isahdia} gives 1 for only two cases
\begin{equation}
\begin{split}
&\text{case 1:} \quad m=m' \text{,} \quad  n=n'\text{,} \quad  l=l', \\
&\text{case 2:} \quad m=n' \text{,} \quad  n=m'\text{,} \quad  l=l'. \\
\label{apx:eq:RV_single_po_case}
\end{split}
\end{equation}
Summing \eqref{apx:eq:isahdia} over both cases \eqref{apx:eq:RV_single_po_case} yields
\begin{equation}
\begin{split}
&G(f) =2\gamma^2\sum_{\forall i}\delta\left(f- f_i\right)e^{2\text{Re}\left[\Gamma\left(z,f\right)\right]} f_0^{3}  \sum_{\forall (m,n,l) \in A_i}\\
&\cdot G_{\text{Tx}}(f_n)G_{\text{Tx}}(f_m)G_{\text{Tx}}( f_l)\left|\mu\left(f_n,f_m,f_l\right)\right|^2\\
&\cdot \left[\frac{1}{2}H^2\left( f_m - f_l\right)+\frac{1}{2}H\left( f_m - f_l\right)H\left( f_{n} - f_{l}\right)\right]
\label{apx:eq:das123da}
\end{split}
\end{equation}
For the non-degenerate set $A_i$, we have that $f_m-f_l+f_n=f_i$ and for a given frequency triplet $\left(f_i,f_m,f_n\right)$ it follows that $f_l=f_m+f_n-f_i$. We can therefore transform the sum in \eqref{apx:eq:das123da} into a two dimensional sum as 
\begin{equation}
\begin{split}
&G(f) =2\gamma^2\sum_{\forall i}\delta\left(f- f_i\right)e^{2\text{Re}\left[\Gamma\left(z,f\right)\right]} f_0^{3}  \sum_{\forall m}\sum_{\forall n}\\
&\cdot G_{\text{Tx}}(f_n)G_{\text{Tx}}(f_m)G_{\text{Tx}}( f_m+f_n-f_i)\left|\mu\left(n,m,m+n-i\right)\right|^2\\
&\cdot \frac{1}{2}\left[ H^2\left( f_i-f_n\right)+H\left( f_i-f_n\right)H\left( f_i-f_m\right)\right]
\label{apx:eq:psd2}
\end{split}
\end{equation}
As in \cite{Semrau_2018_tgn}, we define the normalized signal power profile of a frequency component as $\rho(z,f)=e^{\int_0^z g\left(\zeta,f\right) d\zeta}$ and rewrite \eqref{apx:eq:psd2} as an integral expression by letting $f_0 \to 0$ and relating the NLI to the power level at the transmitter (i.e. multiplying by $\rho(L,f)^{-1}=e^{-2\text{Re}\left[\Gamma\left(z,f\right)\right]}$)
\begin{equation}
\begin{split}
&G(f) =2\gamma^2 \int \int  df_1df_2 \ G_{\text{Tx}}(f_1)G_{\text{Tx}}(f_2)G_{\text{Tx}}(f_1+f_2-f)\\
&\cdot\left|\int_0^Ld\zeta\ \sqrt{\frac{\rho(\zeta,f_1)\rho(\zeta,f_2)\rho(\zeta,f_1+f_2-f)}{\rho(\zeta,f)}}e^{j\phi\left(f_1,f_2,f,\zeta\right)}\right|^2
\\
&\cdot \left[R^2\left( f-f_1\right)+R\left( f-f_1\right)R\left( f-f_2\right)\right],
\label{apx:result_1pol}
\end{split}
\end{equation}
where we defined 
\begin{equation}
\begin{split}
&R(f) = \frac{1}{\sqrt{2}}\Re \left\{H(f)\right\},
\label{eq:dada}
\end{split}
\end{equation}
to include the real part of the complex Raman spectrum. The normalization factor $\frac{1}{\sqrt{2}}$ in \eqref{eq:dada} is introduced to be consistent with the prefactor 2 in  \eqref{apx:result_1pol} and GN model approaches in the absence of a delayed fiber response. Eq. \eqref{apx:result_1pol} is the ISRS GN model for single polarisation extended to account for the real Raman spectrum \eqref{eq:GNmodel_single_pol}.
\subsection{Dual polarized signals}
\label{appendix:sec:dual_pol}
In this section, the result in \eqref{apx:result_1pol} is extended for dual polarized signals. The derivation is similar to the single polarisation case using the generalized Manakov equation \eqref{eq:GME_time}. Inter-channel stimulated Raman scattering is, again, modelled as a frequency and distance dependent gain as in \cite{Semrau_2018_tgn}. The derivation for dual polarized signals is similar to the single polarization case derived in Sec. \ref{appendix:sec:single_pol}, applied to each polarization. However, an essential difference in the dual polarized case is the averaging over the random variables in Eq. \eqref{apx:eq:isahdia}. For the case of dual polarization, the averaging terms are
\begin{equation}
\begin{split}
&\underbrace{\textbf{E}\left[\xi_{\text{x},m}\xi_{\text{x},n}\xi^*_{\text{x},l} \xi^*_{\text{x},m'}\xi^*_{\text{x},n'}\xi_{\text{x},l'}\right]}_{\text{=1, for cases 1,2 \eqref{apx:eq:RV_single_po_case}}}\\
+&\underbrace{\textbf{E}\left[\xi_{\text{x},m}\xi_{\text{x},n}\xi^*_{\text{x},l} \xi^*_{\text{y},m'}\xi^*_{\text{x},n'}\xi_{\text{y},l'}\right]}_{\approx 0}\\
+&\underbrace{\textbf{E}\left[\xi_{\text{y},m}\xi_{\text{x},n}\xi^*_{\text{y},l} \xi^*_{\text{x},m'}\xi^*_{\text{x},n'}\xi_{\text{x},l'}\right]}_{\approx 0}\\
+&\underbrace{\textbf{E}\left[\xi_{\text{y},m}\xi_{\text{x},n}\xi^*_{\text{y},l} \xi^*_{\text{y},m'}\xi^*_{\text{x},n'}\xi_{\text{y},l'}\right]}_{\text{=1, for case 3 \eqref{apx:eq:RV_dual_po_case}}},\\
\end{split}
\end{equation}
where the first case resembles the averaging in Eq. \eqref{apx:eq:isahdia} which is the contribution arising from co-polarization. The second and third average result in negligible contributions (cf. \cite[Sec. IV E]{Poggiolini_2012_ada}). The fourth average, however, yields a cross-polarization contribution for 
\begin{equation}
\begin{split}
&\text{case 3:} \quad m=m' \text{,} \quad  n=n'\text{,} \quad  l=l'. \\
\label{apx:eq:RV_dual_po_case}
\end{split}
\end{equation}
Summing over all three cases, \eqref{apx:eq:RV_single_po_case} and \eqref{apx:eq:RV_dual_po_case}  yields the ISRS GN model extended for the real Raman spectrum as
\begin{equation}
\begin{split}
&G(f) =\frac{16}{27}\gamma^2 \int \int  df_1df_2 \ G_{\text{Tx}}(f_1)G_{\text{Tx}}(f_2)G_{\text{Tx}}(f_1+f_2-f)\\
&\cdot\left|\int_0^Ld\zeta\ \sqrt{\frac{\rho(\zeta,f_1)\rho(\zeta,f_2)\rho(\zeta,f_1+f_2-f)}{\rho(\zeta,f)}}e^{j\phi\left(f_1,f_2,f,\zeta\right)}\right|^2
\\
&\cdot \frac{9^2}{8^23}\left[2H^2\left( f-f_1\right)+H\left( f-f_1\right)H\left( f-f_2\right)\right].
\label{apx:eq:djaisdja}
\end{split}
\end{equation}
The normalization factor $\frac{9^2}{8^23}$ is introduced to be consistent with the prefactor $\frac{16}{27}$ in \eqref{apx:eq:djaisdja} and GN model approaches in the absence of a delayed nonlinear response for dual polarized signals. Defining 
\begin{equation}
\begin{split}
&R(f) =\frac{9}{8\sqrt{3}}\Re \left\{H(f)\right\}
\end{split}
\end{equation}
yields the ISRS GN model extended for the real part of the complex Raman spectrum as in Eqs. \eqref{eq:ISRSGNmodel_rho}\eqref{eq:ISRSGNmodel_Cr}.

\section*{Acknowledgment}
The authors would like to thank Prof. Magnus Karlsson from Chalmers University for helpful discussions. The authors would also like to thank S. Makovejs from Corning for providing the experimental fibre data.

\ifCLASSOPTIONcaptionsoff
  \newpage
\fi

\bibliographystyle{IEEEtran}
\bibliography{IEEEabrv,ref}

\begin{IEEEbiographynophoto}{Daniel Semrau}
(S’16) received the B.Sc. degree in electrical engineering from the Technical University of Berlin, Berlin, Germany, in 2013, the M.Sc. degree in photonic networks engineering from Scuola Superiore Sant’Anna, Pisa, Italy, and Aston University, Birmingham, U.K., in 2015. In 2015, he joined the Optical Networks Group, University College London, U.K., where he received his Ph.D. degree in 2020. In 2018, Daniel was presented with the Graduate Student Fellowship award of the IEEE Photonics Society. His research interests are mainly focused on channel modeling, physical layer-aware optical networking, and ultra-wideband transmission coherent optical communications.
\end{IEEEbiographynophoto}

\begin{IEEEbiographynophoto}{Eric Sillekens}
(S’16) received his BSc and MSc in electrical engineering from the Eindhoven University of Technology in 2012 and 2015 respectively, with his research focussed on advanced coded modulation for optical fibre transmission systems. He is currently a PhD research student in the optical networks group at Univesity College London (UCL) and is supervised by Dr. R. Killey. He is working to holistically optimise long-haul fibre transmission systems, where his interest is in coded modulation and machine learning. He has designed modulation formats with a trade off between fibre nonlinearity and shaping gain.
\end{IEEEbiographynophoto}

\begin{IEEEbiographynophoto}{Robert I. Killey}
(SM’17) received the B.Eng. degree in electronic and communications engineering from the University of Bristol, Bristol, U.K., in 1992, the M.Sc. degree from University College London (UCL), London, U.K., in 1994, and the D.Phil. degree from the University of Oxford, Oxford, U.K., in 1998. He is currently a professor with the Optical Networks Group at UCL. His research interests include nonlinear fiber effects in WDM transmission, advanced modulation formats, and digital signal processing for optical communications. He has participated in many European projects, including ePhoton/ONe, Nobel, BONE and ASTRON, and national projects. He is currently a Principal Investigator in the EPSRC funded UNLOC project. He was with the technical program committees of many international conferences including European Conference on Optical Communication, Optical Fiber Communication Conference ACP, and OECC. He was an Associate Editor of the IEEE/OSA Journal of Optical Communications and Networking and is currently an Associate Editor of the Journal of Lightwave Technology.
\end{IEEEbiographynophoto}

\begin{IEEEbiographynophoto}{Polina Bayvel}
(F'10) received the B.Sc. (Eng.) and Ph.D. degrees in electronic and electrical engineering from UCL (University of London), in 1986 and 1990, respectively. 
In 1990, she was with the Fiber Optics Laboratory, General Physics Institute,
Moscow, Russian Academy of Sciences, under the Royal Society Postdoctoral
Exchange Fellowship. She was a Principal Systems Engineer with STC Sub-
marine Systems, Ltd., London, U.K., and Nortel Networks (Harlow, U.K., and
Ottawa, ON, Canada), where she was involved in the design and planning of
optical fibre transmission networks. During 1994–2004, she held a Royal
Society University Research Fellowship at University College London (UCL),
London, U.K., where she became a Chair in Optical Communications and Net-
works. She is currently the Head of the Optical Networks Group, UCL, which
she set up in 1994. She has authored or coauthored more than 300 refereed jour-
nal  and  conference  papers.  Her  research  interests  include  wavelength-routed
optical networks, high-speed optical transmission, and the study and mitigation
of fibre nonlinearities. She is a Fellow of the Royal Academy of Engineering, IEEE,
the Optical Society of America and the U.K. Institute of Physics. She is Honorary Fellow of the Institution of Engineering and Technology (FIET). She was a recipient the Royal Society Wolfson
Research Merit Award (2007–2012), the 2013 IEEE Photonics Society Engi-
neering  Achievement  Award, the  2014  Royal  Society  Clifford  Patterson
Prize Lecture and Medal and 2015 Royal Academy of Engineering Colin Campbell Mitchell Award. She leads the UK EPSRC Programme TRANSNET (2018-2024).
\end{IEEEbiographynophoto}

\end{document}